\documentclass[a4paper,11pt]{article}

\pdfoutput=1

\usepackage{fullpage}

\usepackage{mathrsfs}
\usepackage{comment}
\usepackage{centernot}
\usepackage{mathtools}
\usepackage{stmaryrd}

\usepackage{latexsym,amssymb,amstext,amsmath}
\usepackage{hyperref}
\usepackage{graphbox,graphicx}
\usepackage{subcaption}
\usepackage{amsthm}
\usepackage{esint}
\usepackage{float}
\usepackage{color} 
\usepackage{tikz}
\usepackage{cite}

\def\be {\begin{equation}}
\def\ee {\end{equation}}
\def\bea {\begin{eqnarray}}
\def\eea {\end{eqnarray}}

\def\beq{\begin{equation}}
\def\eeq{\end{equation}}
\def\beqa{\begin{eqnarray}}
\def\eeqa{\end{eqnarray}}

\theoremstyle{definition}

\DeclareMathOperator{\Tr}{Tr}

\begin{document}

\baselineskip=15pt

\begin{center}

\vspace*{ 0.0cm}
{\Large {\bf Classical integrability in the presence of a cosmological \\
\vskip 2mm 
constant:
analytic and machine learning results}
}
\\[0pt]
\vspace{0.3cm}
\bigskip
{{\bf  Gabriel Lopes Cardoso}},
{{\bf  Dami\'an Mayorga Pe\~na}},
{{\bf Suresh Nampuri}}
\bigskip 

\vspace{0.cm}
{\it \small 
Center for Mathematical Analysis, Geometry and Dynamical Systems,
Department of Mathematics, Instituto Superior T\'ecnico, Universidade de Lisboa,
1049-001 Lisboa, Portugal\\[0.5cm]}

{\small gabriel.lopes.cardoso@tecnico.ulisboa.pt, damian.mayorga.pena@tecnico.ulisboa.pt,
nampuri@gmail.com}
\\[0.4cm]
\end{center}

\begin{center} {\bf Abstract } 
\end{center}
\noindent
We study the integrability of two-dimensional theories that are obtained by a dimensional reduction of certain four-dimensional gravitational theories describing the coupling of Maxwell fields and neutral scalar fields to gravity in the presence of a potential for the neutral scalar fields.
For a certain solution subspace, we demonstrate partial integrability by showing that a subset of the equations of motion in two dimensions are the compatibility conditions for a linear system.
Subsequently, we study the integrability of these two-dimensional models from a complementary one-dimensional point of view,  framed in terms of Liouville integrability.
In this endeavour,  we employ various machine learning techniques to systematise our search for numerical Lax pair matrices for these models, as well as conserved currents expressed as functions of phase space variables.

\vskip 5mm


\vspace{0.4cm}

\section{Introduction}

It has long been established \cite{Belinsky:1971nt,Breitenlohner:1986um,Nicolai:1991tt,Lu:2007jc} that certain 4D gravitational theories 
at the two-derivative level, describing the coupling of Maxwell fields and neutral scalar fields to gravity in the absence of a cosmological constant or, more generally, a potential for the scalar fields, 
when dimensionally reduced to two dimensions, 
yield 2D non-linear sigma-models coupled to gravity, whose equations of motion are classically integrable, i.e. that can be viewed as compatibility equations of a linear system.
This has recently been the subject of revived interest \cite{Katsimpouri:2012ky,Katsimpouri:2013wka,Chakrabarty:2014ora,Camara:2017hez,Cardoso:2017cgi,Aniceto:2019rhg,Penna:2021kua,Camara:2022gvc,Camara:2024ham}.
The dimensional reduction is performed by means of a two-step procedure. In the first step one reduces over an isometry direction down to three dimensions, and then Hodge dualises all Maxwell field strengths into neutral scalar fields.  
We will only consider the reduction of 4D gravitational theories satisfying the condition that 
the resulting target space for the enlarged set of neutral scalar fields is a symmetric space $G/H$. A further reduction along a second commuting isometry direction results in a 2D gravitational model whose field equations are the compatibility equations for a linear system. The latter, introduced by Belinski-Zakharov \cite{Belinsky:1971nt}, has an equivalent description given by
Breitenlohner-Maison \cite{Breitenlohner:1986um,Lu:2007jc}, as shown in \cite{Figueras:2009mc}. 
In this paper we will work with the Breitenlohner-Maison (BM) linear system.

The addition of a cosmological constant or a potential for the neutral scalar fields to these four-dimensional gravitational theories will, in general, not preserve the integrable nature of the dimensionally reduced equations of motion. Aspects of this were discussed in \cite{Leigh:2014dja,Klemm:2015uba}. In this paper we show that in the presence of a scalar potential, 
for a certain solution subspace defined by \eqref{eq:rhot}, a 
subset of the equations of motion in two dimensions can still be viewed as being the compatibility conditions of a linear system.
We further show that this linear system admits a Lax pair description in terms of differential operators \cite{cmn}, and reduces to the  BM system when the scalar potential vanishes.
Hence, this linear system framework for studying integrability only applies to solutions that continue to exist in the absence of a scalar potential. 

Further, these solutions admit a complementary 1D description, thereby enabling us to independently study 
Liouville integrability in 1D expressed in 
terms of Lax pair matrices.
Here, we use analytic and machine learning (ML) techniques to search for Lax pair matrices as well as conserved currents in the 1D setup. ML techniques have been previously implemented in the search for Lax pairs in \cite{Krippendorf:2021lee,Lal:2023dkj}. Here we will adopt a different strategy by framing the Lax pair search as a search for conserved currents. This approach lends itself easily to interpretable results as we show that it admits a symbolic regression implementation. 

Our results can be summarized as follows. For a subclass of 4D gravitational backgrounds we show that in the presence of a potential for the neutral scalar fields, a subset of the 2D
equations of motion can be viewed as compatibility equations for a linear system that reduces to the BM linear system when the potential vanishes.

This subclass of 4D gravitational solutions admits a 1D description.   This enables us to explore the integrability properties of this subclass in the framework of 1D Liouville integrability.

Liouville integrability for a model with a phase space of dimension $n$ demands $n$ independent Poisson commuting conserved currents. These are first integrals of motion,  differentiable functions of phase space variables that are constant on each on-shell trajectory in the solution subspace under consideration and do not take the same value on all such trajectories. One way of encoding such a set of conserved currents is via a pair of Lax matrices $L$ and $M$ satisfying the Lax equation \bea \frac{dL}{dt}-[L,M] =0,\eea with both matrices being non-zero matrix valued functions of phase space variables. In this approach, the existence of these matrices constitutes a necessary but not sufficient condition for Liouville integrability. Sufficiency is achieved by extracting the required number of independent Poisson commuting conserved currents from the Lax pair. 

A given integrable 1D theory lends itself to multiple Lax pair writings of different dimensions, and in order to systematise and simplify our search for these pairs and the consequent identification of conserved currents, we define a `canonical' frame for them where $M$  is  a unit matrix of dimension $n$, while $L$ is a diagonal $n \times n$ matrix with each of the diagonal entries being a conserved function of phase space variables. As we restrict ourself to time-independent models, we can and therefore do choose one of the diagonal elements to be the Hamiltonian. Liouville integrability is then equivalent to the existence of such a Lax matrix where the diagonal elements are linearly independent and Poisson commute with each other. We mount a search for Lax pairs, both canonical and generic, using both analytic and ML techniques. Our findings include ML driven searches for numerical Lax pairs,  analytic results for the Lax pairs and analytic and ML symbolic regression construction of conserved currents. 
The 2D linear system and 1D Liouville integrability approaches make up two complementary and distinct original treatments for this problem.

This paper is organised as follows. In Section \ref{sec:2dr} we write down 2D
non-linear sigma models and the equations of motion resulting from the dimensional reduction of specific 4D gravity theories. We identify a specific subclass of solutions amenable to a modified BM linear system. 
In Section \ref{sec:bmmod} we write down the modified BM linear system as well as Lax pair descriptions in terms of differential operators for the chosen solution subclass, and we give two illustrative examples. In Section \ref{sec:red1d} we discuss the Liouville integrability of the systems from a 1D point of view. In Section \ref{sec:scla1D} we construct analytic Lax pairs in two models. In Section \ref{sec:nums} we set up an ML strategy to determine numerical Lax pairs for the 1D systems studied above. In Section \ref{sec:in} we present
interpretable as well as numerical ML results 
for 
conserved currents in these models.
Finally in Section \ref{sec:conc} we conclude with comments on our findings and comparisons with other ML approaches. In Appendix \ref{sec:eomsL2} we discuss the derivation of the field equations 
in two dimensions.
In Appendix \ref{sec:A} we present families of Lax pair matrices for certain 1D systems. Appendix \ref{sec:MLp} contains further details about our ML experiments.

\section{Dimensional reduction and field equations in two dimensions  \label{sec:2dr}}

We consider a Lagrangian in four space-time dimensions describing the coupling of Maxwell fields and neutral scalar fields $\phi$ to gravity.
We allow for the presence of a potential $V(\phi)$ for the neutral scalar fields. 
We then assume that we can dimensionally reduce this theory down to two dimensions using a well-known 2-step procedure. Here we follow \cite{Lu:2007jc}.

In the first step,
we reduce to three dimensions over an isometry direction $y$ (which is time-like when $\lambda =1$ and
space-like when $\lambda = -1$),
\bea
ds_4^2 = - \lambda \, \Delta \left( dy + B d\varphi \right)^2 + \Delta^{-1} \, ds_3^2 \;\;\;,\;\;\; \lambda = \pm 1.
\label{line4d}
\eea
The metric factors $\Delta$ and $B$ are taken to be independent of $y$.
In three space-time dimensions, we Hodge dualise the Maxwell field strengths into scalar fields. 
We will only consider the reduction of 4D gravitational theories satisfying the condition that 
the resulting target space for the enlarged set of neutral scalar fields is a symmetric space $G/H$.
The resulting
three-dimensional Lagrangian then describes the coupling of these fields 
to three-dimensional gravity, in the presence
of a potential $V(\phi)$. It takes the form \cite{Lu:2007jc} 
\bea
L_3 = \sqrt{| g_3|} \left( R_3 - \frac14 g^{M N} \Tr \left( A_M A_N \right) - \Delta^{-1} \, V(\phi) 
\right)\;,
\label{lag3}
\eea
where the enlarged set of neutral scalar fields is encoded in a matrix $M \in G/H$\footnote{Note that this matrix $M$ appearing in the context of 2D integrability is different from the Lax matrix $M$ that appears in the study of 1D Liouville integrability.}. Here $A \equiv M^{-1} d M$ denotes a matrix 1-form. The symmetric space $G/H$ is endowed with an involutive Lie algebra automorphism $\natural$,
which is used to decompose the Lie algebra of $G$ into the direct sum $ \mathfrak{g} = \mathfrak{h} \oplus \mathfrak{p} $. Then, 
$M \in G/H$ is identified with $M \in G/H \simeq \exp \mathfrak{p}$ and 
satisfies $M = M^{\natural}$. 

In the second step, 
we further reduce the model over a second commuting isometry direction $\varphi$ using
\bea
ds_3^2 = e^{2 \Sigma  } \, ds_2^2 + {\tilde \rho}^2  \,  d\varphi^2 \;. 
\label{3dlin}
\eea
In the above we take ${\tilde \rho}  > 0$. 
We denote the space-time coordinates in two dimensions by $(\rho, v)$, which we take to satisfy $\rho > 0 , v \in \mathbb{R}$. 
The fields $M, \Sigma, {\tilde \rho}$ are functions of $(\rho, v)$ only. 
Taking $ds_2^2$ to be a flat metric given by
 \bea
 ds_2^2 = \lambda \, d \rho^2 + dv^2 \;,
 \label{flat2dm}
 \eea
and combining \eqref{3dlin} with \eqref{line4d} results in a four-dimensional line element given by
\bea
\label{eq:lineelem2}
ds_4^2 = - \lambda \, \Delta \left( dy + B d\varphi \right)^2 + \Delta^{-1} \, \left( e^{2 \Sigma}  \left( \lambda \, d \rho^2 + dv^2 \right) 
+ {\tilde \rho}^2  \,  d\varphi^2 \right)
\; \;\;\;,\;\;\; \lambda = \pm 1 .
\label{4dline}
\eea

The resulting two-dimensional Lagrangian takes the form given in 
\cite{Lu:2007jc}, augmented by the presence of a potential 
\bea
{\cal V} ({\tilde \rho}, \Sigma, \Delta, \phi) \equiv  {\tilde \rho} \, e^{2 \Sigma  } \, \Delta^{-1} \, V(\phi), 
\label{calvpot}
 \eea
 namely
\bea
L_{2} = \sqrt{|g_2|} \, \left[ {\tilde \rho}  \left( R_2 - \frac14 g^{\mu \nu} \Tr \left( A_{\mu} A_{\nu} \right) \right) + 2 g^{\mu \nu} \partial_{\mu} {\tilde \rho} \partial_{\nu} \Sigma - 
{\cal V} ({\tilde \rho}, \Sigma, \Delta, \phi)
\right], 
\label{Lag2d}
\eea
 where 
 \bea
 A_{\mu} = M^{-1} \partial_{\mu} M \;\;\;,\;\;\; \mu = \rho, v \:.
 \eea
Note that ${\cal V}$ depends linearly on $ {\tilde \rho} $. The potential ${\cal V}$ depends on the fields ${\tilde \rho}, \Sigma, \Delta, \phi$, but only the fields
$\Delta, \phi$ (and also $\tilde{B}$, which is defined below in \eqref{btb})  are encoded in the matrix $M$. 

We introduce the Hodge star operator $\star$ in two dimensions, satisfying
\bea
\star d \rho = - \lambda \, dv \;\;\;,\;\;\; \star dv = d \rho \;\;\;,\;\;\; (\star)^2 = - \lambda \, {\rm id} \;.
\label{stho}
\eea
Using $\star$, the field $\tilde B$ is defined in terms of the field $B$ in \eqref{line4d} by \cite{Breitenlohner:1986um}
\bea
\rho \star d {\tilde B} = \Delta^2 \, dB \;.
\label{btb}
\eea

The equations of motion that follow from the two-dimensional Lagrangian \eqref{Lag2d} can be obtained using standard techniques. We present the details of their derivation in Appendix \ref{sec:eomsL2}, for the benefit of the readers.
We now summarise the outcome of this analysis. 
The independent field equations are
\bea
d \left( \tilde{\rho} \star A \right) = dv \wedge d\rho  \, {\tilde \rho} \, G \;\;\;,\;\;\; A = M^{-1} \, d M \;,
\label{fieldA}
\eea
where the combination $ {\tilde \rho} \, G = \nabla^{\mu} \left( {\tilde \rho}  A_{\mu} \right) $ is given by \eqref{rR}, 
 and 
\bea
&& \lambda \partial_{\rho}^2 {\tilde \rho} + \partial_{v}^2 {\tilde \rho} = - {\cal V} , \nonumber\\
&&  \lambda \partial_{\rho}^2 (2 \Sigma )  + \partial_{v}^2  (2 \Sigma ) = 
 -  \frac14 \Tr \left( \lambda \, A_{\rho} A_{\rho } + A_v A_v \right)   -  \frac{{\cal V} }{\tilde \rho} , \nonumber\\ 
  &&
 \partial_{\rho} {\tilde \rho} \, \partial_{\rho} (2 \Sigma) - \lambda  \partial_{v} {\tilde \rho} \, \partial_{v} (2 \Sigma) + 2 \lambda 
  \partial_v^2  {\tilde \rho} 
- \frac14 \, {\tilde \rho}  \,   \Tr \left[ A_{\rho} A_{\rho} - \lambda A_v A_v \right] + \lambda \, {\cal V} = 0 , \nonumber\\
&&
 \partial_{\rho} {\tilde \rho} \, \partial_{v} (2 \Sigma) + \partial_{v} {\tilde \rho} \, \partial_{\rho} (2 \Sigma) - 2 \partial_{\rho} \partial_v 
{\tilde \rho} - \frac12  \, {\tilde \rho}  \,   \Tr \left[ A_{\rho} A_{v} \right] = 0 \;.
\label{eomsdist}
\eea

 \subsection{Subspace of solutions \label{sec:subsol}}

In the following 
we will focus on solutions to the field equations that have the following dependence on $\rho, v$,
\bea\label{eq:rhot}
\tilde{\cal \rho} (\rho,v) = h(\rho) \, g(v) \;\;\;,\;\;\; \Sigma = \Sigma (\rho) \;\;\;,\;\;\; \Phi^I = \Phi^I (\rho)\,.
\eea
In the above $\Phi^I$ denote the scalar fields encoded in $M$, which include $\Delta, {\tilde B}, \phi$, where $\phi$ refers to the set of scalar fields in four dimensions.
This implies
\bea
A_v = M^{-1} \partial_v M = 0 \;.
\eea
Then, the field equation \eqref{fieldA} becomes
\bea
\label{P1}
\lambda \, \partial_{\rho} A_{\rho} +  \lambda \, 
(\partial_{\rho} \log h) \, A_{\rho}
 =
 G \,,
\eea
with ${\tilde \rho} \, G$ given by \eqref{rR}. Note that $G$ is a function of $\rho$ only.
Furthermore, we assume that
\bea\label{eq:alpha}
\partial_v^2 g(v) = -\alpha \, g(v) \;\;\;,\;\;\; \alpha = -1, 0, 1 \;.
\eea
Then, the first  equation in \eqref{eomsdist} becomes
\bea
\lambda \frac{\partial^2_{\rho} h}{h} - \alpha = -
\frac{ \cal V}{\tilde{\cal \rho}}  \;,
\label{h1}
\eea
while the second and third equations become
\bea
&& \lambda \partial_{\rho}^2 (2 \Sigma )  = 
 -  \frac{\lambda}{4} \Tr \left(  A_{\rho} A_{\rho }  \right)   - \frac{{\cal V} }{\tilde \rho} \,, \label{h00}\\
 && \frac{\partial_{\rho} h}{h}  \, \partial_{\rho} (2 \Sigma)  - 2 \lambda \alpha
- \frac14  \,   \Tr \left[ A_{\rho} A_{\rho} \right] + \lambda \,  \frac{{\cal V} }{\tilde \rho}  = 0 .
\label{h22}
 \eea
Combining \eqref{h1} with \eqref{h00} and \eqref{h22} gives
\bea
\partial_{\rho} \left( h \, \partial_{\rho} (2 \Sigma) - 2 \partial_{\rho} h  \right) = 0 \;,
\eea
which is solved by
\bea
2 \Sigma = 2 \log h - c \int \frac{d \rho}{h} \;, \; c \in \mathbb{R}.
\label{sighrel}
\eea
The fourth equation in \eqref{eomsdist} becomes
\bea
h \, \partial_v g \, \partial_{\rho} \log \left( h^{-2}\, e^{2 \Sigma}
\right) = 0 \;.
\eea
When $\partial_v g =0$, this does not impose any further condition on \eqref{sighrel}. On the other hand, when
$\partial_v g \neq 0$, we conclude that
\bea
e^{2 \Sigma} =  h^2 \;,
\label{Srt}
\eea
where we have adjusted the integration constant so as to ensure compatibility with \eqref{sighrel}. Thus, $2\Sigma$ is determined
by \eqref{sighrel}, with $c=0$ when $\partial_v g \neq 0$.

Summarizing, 
the field equations for $\tilde \rho$ and for the scalar fields encoded in $M$ 
are \eqref{P1}, \eqref{h1} and \eqref{h22}, with 
$2\Sigma$
determined in terms of $h$ by \eqref{sighrel}.

When ${\cal V} = 0$, the field equation \eqref{P1} is solved by
\bea
A_{\rho} = \frac{C}{h} \,,
\eea
where $C$ denotes a constant matrix. When $\alpha = \pm 1$ we have
$2 \Sigma = 2 \log h$, and \eqref{h22} becomes
\bea
2 (\partial_{\rho} h)^2 - 2 \lambda \alpha h^2 - \frac14 \, \Tr C^ 2 = 0 \,.
\label{Chpm1}
\eea
Writing the general solution of \eqref{h1} as a linear combination
of two linearly independent basis vectors $h_1 $ and $h_2$ that
satisfy $h_1^2 + \lambda \alpha \, h_2^2 = 1$,
\bea
h = a_1 \, h_1 + a_2 \, h_2 \quad , \quad a_1, a_2 \in \mathbb{R},
\eea
we find that \eqref{Chpm1} imposes the following condition on the constant matrix $C$,
\bea
2 \left( a_1^2 - \lambda \alpha \, a_2^2 \right) = \frac14 \, \Tr C^ 2 \;.
\eea
On the other hand, when $\alpha = 0$, 
\eqref{h22} becomes
\bea
2 (\partial_{\rho} h)^2 - c\, \partial_{\rho} h - \frac14 \, \Tr C^2 = 0 \,.
\label{Chpm2}
\eea
Writing the general solution of \eqref{h1} as
\bea
h = a_1 + a_2 \, \rho \, \quad a_1, a_2 \in \mathbb{R},
\eea
we obtain the following condition on the constant matrix $C$,
\bea
2 a_2^2 - c\, a_2 - \frac14 \, \Tr C^2 = 0 \,.
\eea

In the next section we will introduce a linear system in two dimensions whose solvability condition is the field equation \eqref{P1}.
To do so, we will write $\tilde \rho$ as a sum of two terms, ${\tilde \rho} = {\tilde \rho}_0 + {\tilde \rho}_P$, where
\bea
\lambda \partial_{\rho}^2 {\tilde \rho}_0 + \partial_{v}^2 {\tilde \rho}_0 = 0 \;\;\;,\;\;\;  \lambda \partial_{\rho}^2 {\tilde \rho}_P + \partial_{v}^2 {\tilde \rho}_P = - {\cal V} \,.
\eea
We then introduce the field ${\tilde v}_0$ by
\bea
\star d {\tilde \rho}_0 = - \lambda \, d {\tilde v}_0 \;\;\;,\;\;\; 
\star d {\tilde v}_0 = d {\tilde \rho}_0 .
\label{rvcoord}
\eea
We assume that ${\tilde \rho}_0$ has the form
\bea
\tilde{\cal \rho}_0 (\rho,v) = h_0(\rho) \, g_0(v) \,.
\label{tildr0rv}
\eea
The spectral parameter entering the linear system will be defined in terms of $({\tilde \rho}_0, {\tilde v}_0)$.

\section{Integrability in two dimensions
\label{sec:bmmod}}

We now turn to the question of integrability of the field equations in two dimensions.
In the absence of a potential ${V}$, it is well known that the field equations for $\tilde \rho$ and $M$ constitute the compatibility conditions for the Breitenlohner-Maison linear system \cite{Breitenlohner:1986um,Lu:2007jc}.
In the presence of a potential ${V}$, no such linear system has been written down. 
In the following section, we write down a linear system 
for the subspace of solutions discussed in Section \ref{sec:subsol}, such that the corresponding field equation \eqref{P1} for $M$ is the compatibility
condition for this linear system, as we show below.

We proceed by first reviewing the Breitenlohner-Maison (BM) linear system for $V(\phi) = 0$ in the language of Lax pairs \cite{cmn}
generalising the discussion given there. We then proceed to construct the linear system for $V(\phi) \neq 0$.

\subsection{ $V (\phi) = 0$: A Lax pair for the BM linear system}

In the absence of a potential $ V$, the field equations for $M$ and $\tilde \rho$ are given by (cf. \eqref{fieldA} and \eqref{eomsdist})
 \bea
 \label{feqV0}
 F &=& dA + A \wedge A = 0 \Longleftrightarrow F_{\rho v}= 0 \;, \nonumber\\
 d \left( {\tilde \rho} \star A \right) &=& 0 \Longleftrightarrow  \lambda \partial_{\rho} A_{\rho}+ 
 \partial_v A_v + \frac{\lambda}{\tilde \rho} \,
 \frac{\partial {\tilde \rho}}{ \partial \rho} \, A_{\rho} + \frac{1}{\tilde \rho}\,  \frac{\partial {\tilde \rho}}{\partial v} \, A_v = 0 \;, \nonumber\\
 d \star d {\tilde \rho} &=& 0 \Longleftrightarrow 
 \lambda \partial_{\rho}^2 {\tilde \rho} + \partial_{v}^2 {\tilde \rho} = 0 \,.
 \eea
It is well known that these field equations are the compatibility conditions for an auxiliary system of first-order differential equations, 
called Breitenlohner-Maison linear system 
\cite{Breitenlohner:1986um}. This linear system (a Lax pair) takes the form \cite{Lu:2007jc}
\bea
\tau \left( d + A \right) X = \star \, d X \,,
\label{BM}
\eea
where $X$ is a matrix function, and 
where $\tau$ is a function taken from the set of functions $\varphi_{\omega}(\rho,v)$ of the form \cite{Aniceto:2019rhg}
\bea
\label{settau}
\frac{- \lambda (\omega -{\tilde v} ) \pm \sqrt{ (\omega - 
{\tilde v} )^ 2 + \lambda {\tilde \rho}^2}}{\tilde \rho} \;\;\;,\;\;\; \omega \in \mathbb{C} 
\;,
\eea
with $\tilde v$ defined as in \eqref{rvcoord}. Note that 
 \bea
 \star d{\tilde \rho} = - \lambda \, d {\tilde v} \;\;\;,\;\;\; \star d {\tilde v} = d {\tilde \rho} \qquad \Longleftrightarrow \qquad \frac{\partial {\tilde \rho}}{\partial \rho} = 
 \frac{\partial {\tilde v}}{\partial v} \;\;\;,\;\;\; \frac{\partial {\tilde \rho}}{\partial v} = - \lambda 
 \frac{\partial {\tilde v}}{\partial \rho} 
 \label{dilddil2}
 \eea
as well as
 \bea
 \frac{\partial \tau}{\partial {\tilde \rho}} = \frac{ \tau (\lambda - \tau^2)}{{\tilde \rho} (\lambda + \tau^2)} \quad, \quad 
 \frac{\partial \tau}{\partial {\tilde v}} = \frac{ 2 \lambda  \tau^2 }{{\tilde \rho} (\lambda + \tau^2)} ,
 \eea
 and hence
 \bea
 -  \frac{\partial \tau}{\partial {\tilde v}} + 
 \tau  \frac{\partial \tau}{\partial {\tilde \rho}} = - \frac{\tau^2}{\tilde \rho} \quad , \quad 
 \lambda  \frac{\partial \tau}{\partial {\tilde \rho}} + \tau  \frac{\partial \tau}{\partial {\tilde v}} = \lambda  \frac{\tau}{\tilde \rho} .
 \label{reltrel}
 \eea

The linear system \eqref{BM} 
can be expressed in terms of a pair of differential operators 
$({\cal  L},  {\cal  M})$ \cite{cmn},
\bea
\tau \left( d + A \right) X = \star \, d X 
\Longleftrightarrow  {\cal  L} X =  {\cal  M} X =0 \;,
\eea
where
   \bea
 {\cal  L} &=& - \partial_v + \tau \, D_{\rho} , \nonumber\\
 {\cal M} &=& \lambda \, \partial_{\rho} + \tau \, D_v \;\;\;,\;\;\; \lambda = \pm 1, 
 \label{LML}
 \eea
 with
 \bea
 D_{\mu} = \partial_{\mu} + A_{\mu} \;\;\;,\;\;\; \mu= \rho, v \;.
 \eea
For the case when $\tilde \rho$ is identified with the coordinate $\rho$, it was shown in \cite{cmn} that the compatibility condition $0 = [{\cal L}, {\cal M}] X$ implies the field equations \eqref{feqV0}. Here we generalise the discussion given in \cite{cmn} to the case when $\tilde \rho$ is a function of $(\rho,v)$. 
We compute the compatibility condition $0 = [{\cal L}, {\cal M}] X$,
\bea
0 = [{\cal L}, {\cal M}] X= - \frac{\lambda}{\tau} \left( - \partial_v \tau + \tau \partial_{\rho} \tau \right) \partial_{\rho}
- \left( \lambda \partial_{\rho} \tau + \tau \partial_v \tau \right) D_{\rho} -\tau \left( \partial_v A_v + \lambda \partial_{\rho} A_{ \rho} \right) + \tau^2 \, F_{\rho v},
\label{commr1}
\eea
 where we used ${\cal M} X = 0$ once. Next, using \eqref{dilddil2}  as well as \eqref{reltrel}, we establish
 \bea
 - \partial_v \tau + \tau \partial_{\rho} \tau &=& - \frac{\tau^2}{\tilde \rho} \frac{ \partial {\tilde \rho}}{\partial \rho} + \lambda \frac{\tau}{\tilde \rho} \frac{ \partial {\tilde v}}{\partial \rho} \;,\nonumber\\
  \lambda \partial_{\rho}\tau + \tau \partial_v \tau &=& \lambda \frac{\tau}{\tilde \rho} \frac{ \partial {\tilde \rho}}{\partial \rho} - 
\frac{\tau^2}{\tilde \rho} \frac{ \partial {\tilde \rho}}{\partial v} .
 \eea
 Inserting this into \eqref{commr1} gives
 \bea
0 &=& [{\cal L}, {\cal M}] X= \frac{1}{\tilde \rho} \frac{  \partial {\tilde \rho}}{\partial v} \left(\tau A_v +  \lambda \partial_{\rho} + \tau^2 D_{\rho} \right) X
 \nonumber\\
&&
-\tau \left( \partial_v A_v + \lambda \partial_{\rho} A_{ \rho}  + \frac{\lambda}{\tilde \rho} \,
 \frac{\partial {\tilde \rho}}{ \partial \rho} \, A_{\rho} + \frac{1}{\tilde \rho} \,  \frac{\partial {\tilde \rho}}{\partial v} \, A_v\right) X + \tau^2 \, F_{\rho v} X.
\label{commr2}
\eea
Next, using both ${\cal L} X = 0$ and ${\cal M} X = 0$  we infer the relation
\bea
\left(  \lambda \partial_{\rho} + \tau^2 D_{\rho} \right) X = \left( - \tau D_v + \tau \partial_v \right) X = - \tau A_v X , 
\eea
 and hence the first line in \eqref{commr2} cancels out, so that
  \bea
0 = [{\cal L}, {\cal M}] X =\left[  -\tau \left( \partial_v A_v + \lambda \partial_{\rho} A_{ \rho}  + \frac{\lambda}{\tilde \rho} \,
 \frac{\partial {\tilde \rho}}{ \partial \rho} \, A_{\rho} + \frac{1}{\tilde \rho} \,  \frac{\partial {\tilde \rho}}{\partial v} \, A_v\right)  + \tau^2 \, F_{\rho v} \right] X.
\eea
Assuming that the matrix $X$ is invertible, we obtain
\bea
 \partial_v A_v + \lambda \partial_{\rho} A_{ \rho}  + \frac{\lambda}{\tilde \rho} \,
 \frac{\partial {\tilde \rho}}{ \partial \rho} \, A_{\rho} + \frac{1}{\tilde \rho} \,  \frac{\partial {\tilde \rho}}{\partial v} \, A_v  = 0  \quad \wedge \quad    F_{\rho v} = 0,
\label{eomFl}
\eea
which, together with \eqref{dilddil2}, are the field equations \eqref{feqV0}.  Finally,  $2 \Sigma$ 
is determined by integrating the last two equations of \eqref{eomsdist}. Their solvability is a consequence of the first two equations in \eqref{eomsdist}.

When ${V} \neq 0$, the field equations for $M$ and ${\tilde \rho}$ are modified, and in general these modified field equations will no longer be the compatibility conditions for a linear system.  Below we will show that when restricting to the subspace of solutions discussed in Section \ref{sec:subsol}, one can write down a linear system whose
compatibility
condition is  the modified field equation \eqref{P1} for $M$.
Since the field equation for $\tilde \rho$
seizes to be of the form given in \eqref{feqV0}, this linear system cannot be formulated in terms of $\tilde \rho$, but instead
will have to be formulated in terms of a pair 
$({\tilde \rho}_0, {\tilde v}_0)$ that satisfies \eqref{rvcoord}.
The field equation for $\tilde \rho$ given in \eqref{h1}
will therefore not arise as a
compatibility condition for the linear system and will
have to be imposed separately. Likewise, \eqref{h22} will also have to be verified separately, where we recall that $2 \Sigma$ is given by \eqref{sighrel}.

\subsection{$V (\phi) \neq 0$: A Lax pair \label{sec:modLP}}

We focus on the solution space described in Section \ref{sec:subsol}, in which case $A_v = 0$, and we consider the following modification of 
the Lax pair \eqref{LML} which is induced by a non-vanishing
${\cal V}$,
 \bea
 {\mathcal L} & = &  - \partial_v + \tau \, D_{\rho}  =
 - \partial_v + \tau \left( \partial_{\rho}  + A_{\rho}  \right) \, , \nonumber\\
{\mathcal M} &=& \lambda \, \partial_{\rho} + \tau \partial_v  + \Omega (\rho) \;\;\;,\;\;\; \lambda = \pm 1, 
 \label{LML2}
\eea
where $\tau$ is a function taken from the set of functions \eqref{settau} with $({\tilde \rho},{\tilde v})$ replaced by
$({\tilde \rho}_0,{\tilde v}_0)$, and where ${\tilde \rho}_0$ has the form given in \eqref{tildr0rv}. 
Here, $\Omega (\rho)$ describes the modification of the Lax pair \eqref{LML} due to the presence of ${\cal V}$.

We impose ${\mathcal L} X = {\mathcal M} X = 0$ and study the 
compatibility condition 
\bea
{\mathcal L} X = 0 \; \wedge \; {\mathcal M} X = 0 \Longrightarrow [ {\mathcal L}, {\mathcal M}] X = 0.
\eea
We obtain
\bea
0 = [ {\mathcal L}, {\mathcal M}] X = \left[ \tau \left( \partial_{\rho} \Omega - \lambda \, \partial_{\rho} A_{\rho}  \right) 
  + \left( - \partial_v \tau + \tau \partial_{\rho} \tau  \right) \partial_v - \left( \lambda \, \partial_{\rho} \tau + \tau \partial_v \tau \right) D_{\rho}  \right] X .
\eea
Using $ {\mathcal M} X = 0$
we obtain
\bea
0 = [{\cal L}, {\cal M}] X= \left[\tau \left( \partial_{\rho} \Omega -
\lambda \, \partial_{\rho} A_{\rho}  \right)  - \frac{1}{\tau} \left( - \partial_v \tau + \tau \partial_{\rho} \tau \right) 
\left( \lambda \, \partial_{\rho} + \Omega \right) 
- \left( \lambda \,  \partial_{\rho}\tau + \tau \partial_v \tau \right) D_{\rho} \right] X \;.
\label{commr3}
\eea
Next, using \eqref{dilddil2}  as well as \eqref{reltrel}
with $({\tilde \rho},{\tilde v})$ replaced by
$({\tilde \rho}_0,{\tilde v}_0)$,
we establish
 \bea
 - \partial_v \tau + \tau \partial_{\rho} \tau &=& - \frac{\tau^2}{{\tilde \rho}_0} \frac{ \partial {\tilde \rho}_0}{\partial \rho} + \lambda \frac{\tau}{{\tilde \rho}_0} \frac{ \partial {\tilde v}_0}{\partial \rho} 
 =  - \frac{\tau^2}{{\tilde \rho}_0} \frac{ \partial {\tilde \rho}_0}{\partial \rho} -  
  \frac{\tau}{{\tilde \rho}_0} \frac{ \partial {\tilde \rho}_0}{\partial v} 
 \nonumber\\
  \lambda \, \partial_{\rho}\tau + \tau \partial_v \tau &=& \lambda \frac{\tau}{{\tilde \rho}_0} \frac{ \partial {\tilde \rho}_0}{\partial \rho} - 
\frac{\tau^2}{{\tilde \rho}_0} \frac{ \partial {\tilde \rho}_0}{\partial v} .
\eea
 Inserting this into \eqref{commr3} gives
 \bea
0 &=& [{\cal L}, {\cal M}] X= \tau \left( \partial_{\rho} \Omega - \lambda \, \partial_{\rho} A_{\rho} - \frac{\lambda}{{\tilde \rho}_0} \,
 \frac{\partial {\tilde \rho}_0}{ \partial \rho} \, A_{\rho}  \right) X + 
\frac{1}{{\tilde \rho}_0} \frac{  \partial {\tilde \rho}_0}{\partial v} \left( \lambda \,   \partial_{\rho} + \tau^2 D_{\rho} \right) X
 \nonumber\\
 && + \left(  \frac{\tau}{{\tilde \rho}_0} \frac{ \partial {\tilde \rho}_0}{\partial \rho} +
  \frac{1}{{\tilde \rho}_0} \frac{ \partial {\tilde \rho}_0}{\partial v} \right) \Omega \, X \;.
  \eea
Next, using both ${\cal L} X = 0$ and ${\cal M} X = 0$  we obtain
\bea
\left( \lambda \,  \partial_{\rho} + \tau^2 D_{\rho} \right) X = - \Omega \, X \;,
\eea
 and hence we get
   \bea
0 &=& [{\cal L}, {\cal M}] X= \tau \left[
\left(  \partial_{\rho}  + \frac{1}{{\tilde \rho}_0} \,
 \frac{\partial {\tilde \rho}_0}{ \partial \rho} \right) ( \Omega - \lambda \, A_{\rho} ) \right] X \nonumber\\
 &=& \frac{\tau}{h_0} \, \partial_{\rho}
 \left[ h_0 \,  ( \Omega - \lambda \, A_{\rho} )  \right] X 
 \,.
 \eea
Assuming that the matrix $X$ is invertible, we infer
\bea
 \Omega = \lambda \, \left( A_{\rho} - \frac{C}{h_0} \right)\;,
 \label{Arhofrelund}
 \eea
where $C$ is a constant matrix. 

Demanding that when ${\cal V} =0$ the modified Lax pair reduces to the original Lax pair \eqref{LML}, implies that $\Omega =0$ when switching off
the potential ${\cal V}$. We assume that the solution for $ A_{\rho}$ in the presence of  ${\cal V} \neq 0$ reduces to the one when  ${\cal V} =0$,
and that likewise $h$ reduces to $h_0$. It follows that when 
${\cal V} =0$, $ A_{\rho}$ takes the form
\bea
  A_{\rho}\vert_{{\cal V} =0 } = \frac{C}{h_0} \;.
  \label{AV0}
  \eea
Then, the modification $\Omega$ can be expressed as
\bea
 \Omega = \lambda \, \left( A_{\rho} - A_{\rho}\vert_{{\cal V} =0}  \right) 
 \;.
 \label{deff1}
 \eea
 
To relate $\Omega$ to the potential ${\cal V}$, we 
return to the field equation for $M$ given in \eqref{P1},
which we write as
\bea
\partial_{\rho} \left( h \, A_{\rho} \right) = \lambda \, h \, G \;.
\eea
Both sides of this equation only depend on $\rho$. Integrating over 
$\rho$ gives
\bea
A_{\rho} = \frac{\lambda}{h} \int h \, G \, d \rho + \frac{C}{h} 
\;,
\eea
where we have adjusted the integration constant so as to obtain 
$A_{\rho}\vert_{{\cal V} =0}$ when $G =0$.
Then, $\Omega$ can be expressed as
\bea
 \Omega = \frac{1}{h} \int h \, G \, d \rho + \lambda \left( \frac{h_0}{h} - 1 \right) A_{\rho}\vert_{{\cal V} =0 } 
  \;.
 \eea

We note that the modified Lax pair \eqref{LML2} can also be rewritten as
\bea
\tau \left( d + A \right) X = \star d X - (\Omega \,  dv) X = \star 
\left( d + P \right) X \;\;\;,\;\;\; P = \lambda \, \Omega \, d \rho \;.
\eea
At this state it is important to point out that we write the modified Lax pair only for configurations for which we can define a $V = 0$ limit, as we show in the examples below.

\subsection{Examples}
When dimensionally reducing General Relativity to two dimensions using the 2-step procedure mentioned above, the resulting coset $G/H$ is $SL(2, \mathbb{R})/ SO(2)$. The matrix $M$ is given by \cite{Breitenlohner:1986um}
\bea
M= 
\begin{pmatrix}
\Delta + {\tilde B}^2/\Delta & {\tilde B}/\Delta \\
{\tilde B}/\Delta & 1/\Delta
\end{pmatrix}, 
\eea
where $\Delta$ is the metric factor in \eqref{4dline}, and ${\tilde B}$
is related to the metric factor $B$ by \eqref{btb}.

In the presence of a cosmological constant $V = 2\Lambda_4 = - 6/L^2$,
${\cal V}$ takes the form
\bea
{\cal V} = {\tilde \rho} \, e^{2 \Sigma} \, \Delta^{-1} \, 2 \Lambda_4
\;.
\eea
We now discuss two illustrative examples that arise in General Relativity in four dimensions in the presence of a cosmological constant,
namely
the $AdS_4$-Schwarzschild solution and the $AdS_4$ solution.

\subsubsection{$AdS_4$-Schwarzschild solution \label{sec:schwarzads}}
We consider the $AdS_4$-Schwarzschild solution in General Relativity in the presence of a cosmological constant $V = 2\Lambda_4 = - 6/L^2$. In spherical coordinates, its line element reads
\bea
ds_4^2 = - f(r) \, dt^2 + \frac{1}{f(r)} \, dr^2 + r^2 \left( d \theta^2 + \sin^2 \theta \, d \varphi^2 \right)
\label{lineadsbh}
\eea
with
\bea
f(r) =  1 - \frac{2m}{r} + \frac{r^2}{L^2} > 0 \;.
\label{fmL}
\eea
Denoting the real zero of $f(r)$ by $r_0$, we restrict to $r >  c> r_0$ to ensure that $f(r) > 0$.
The line element can be brought to the form \eqref{4dline} (with $\lambda =1, B= 0$)
by the coordinate transformation
\bea
    \rho (r)=\int_{c}^r\frac{d\tilde{r}}{\sqrt{\tilde{r}^2-2m \tilde{r}+\frac{\tilde{r}^4}{L^2}}} > 0
    \quad ,\quad v=\theta \;,
    \label{rhorrel}
\eea
which satisfies $\rho(c)=0$.
We refrain from giving the explicit expression for $\rho (r)$, which involves an elliptic integral of the first kind. The quantities
$\Delta, \Sigma$ and $\tilde \rho$ are given by
 \bea
\Delta (\rho) = f(r)  \quad , \quad e^{\Sigma (\rho) } = h(\rho) \quad, \quad {\tilde \rho} (\rho, v) = h(\rho) \,
g(v)
\label{dsrt}
\eea
with
\bea
h(\rho) = r \, \sqrt{f(r)}
\quad , \quad g(v) = \sin v \;.
\eea
Then
\bea
A_{\rho} =  \Delta^{-1} \partial_{\rho} \Delta \, \begin{pmatrix}
1 & 0 \\
0 & -1
\end{pmatrix} =\frac{2(L^2 m+r^3 (\rho))}{L^2 h(\rho)} \begin{pmatrix}
1 & 0 \\
0 & -1  \\
\end{pmatrix} \,.
\label{arhoL}
\eea
Let us consider the case when $m \neq 0$. When switching off the
cosmological constant (by sending $L \rightarrow + \infty$), we obtain from \eqref{rhorrel},
\bea
r (\rho) = m \left( 1 + \cosh \left( \rho + \rho_0 \right)\right) \;\;\;,\;\;\; \rho > 0 \;,
\label{rmrho}
\eea
where the constant $\rho_0$ is given by
\bea
\rho_0 = \cosh^{-1} \left( \frac{c}{m} - 1 \right) \;.
\eea
Note that $c/m  - 1 >  1$, since $c> r_0 = 2m$.
Inserting this into \eqref{dsrt} results in
\bea
\Delta_0 &=& \frac{-1+{\rm cosh}\left( \rho + \rho_0 \right)}{1+{\rm cosh}\left( \rho + \rho_0 \right)}\quad , \quad e^{\Sigma_0} = m \sinh \left( \rho + \rho_0 \right) \;, \nonumber\\
{\tilde \rho}_0 &=&
m \sinh \left( \rho + \rho_0 \right) \, \sin v \quad , \quad {\tilde v}_0 = - m\cosh \left( \rho + \rho_0 \right) \, \cos v \;.
\eea
Note that ${\tilde \rho}_0$ is of the form \eqref{tildr0rv} with
\bea
h_0 (\rho) = m\sinh \left( \rho + \rho_0 \right) \quad , \quad g_0 (v) = \sin v \,.
\eea
Using \eqref{rmrho} in \eqref{arhoL} results in
\bea
A_{\rho}\vert_{{\cal V} =0 } =
\frac{2}{\sinh \left( \rho + \rho_0 \right)} \begin{pmatrix}
1 & 0 \\
0 & -1  \\
\end{pmatrix} \,.
\eea
Hence,
the modification $\Omega$ given in
\eqref{deff1}, when evaluated on the solution, reads
\bea
\Omega =  A_\rho - \frac{2}{\sinh \left( \rho + \rho_0 \right)} 
\begin{pmatrix}
1 & 0 \\
0 & -1  \\
\end{pmatrix} \,.
\eea
In order to recover $AdS_4$ space-time we send $m$ to zero. In this limit $A_{\rho}\vert_{{\cal V} =0 }$ shrinks to zero since $\rho_0 \rightarrow + \infty$. We now proceed to analyse the $AdS_4$ solution.

\subsubsection{$AdS_4$ solution}

Now let us set $m=0$ in \eqref{fmL}, in which case the line element \eqref{lineadsbh} describes
an $AdS_4$ space-time in global coordinates. We restrict to the patch $r > c > 0 $.
{From} \eqref{rhorrel} we obtain
\bea
\rho (r) = \log \left( 1 + \sqrt{ 1 + \frac{c^2}{L^2}} \right) + \log \frac{r}{c} - \log \left( 1 + \sqrt{ 1 + \frac{r^2}{L^2}} \right)  \;,
\label{rrho}
\eea
where we have adjusted the integration constant so as to ensure that $\rho (c) =0 $.
Using \eqref{arhoL}, we obtain
\bea
A_{\rho} = 2 \frac{ r^2 (\rho)}{L^2 \sqrt{1 + \frac{r^2 (\rho)}{L^2 }}} \; \begin{pmatrix}
1 & 0 \\
0 & -1  \\
\end{pmatrix} \,,
\label{arhoL2}
\eea
with $r(\rho)$ determined from \eqref{rrho}.
Then, in the limit $L\rightarrow + \infty$ we  have $\rho = \log \frac{r}{c} $, and we obtain
\bea
\Delta_0 = 1 \quad , \quad e^{\Sigma_0} = c \, e^{ \rho} \;\;\;,\;\;\; {\tilde \rho}_0= c \,  e^{ \rho}
 \, \sin v \quad , \quad {\tilde v}_0 = - c \, e^{\rho} \, \cos v \;,
\eea
as well as $A_{\rho}\vert_{{\cal V} =0 } =0 $. Hence, the modification $\Omega$ given in
\eqref{deff1}, when evaluated on the solution, reads
\bea
\Omega = A_{\rho} \,.
\eea

\section{Reduction to one dimension \label{sec:red1d} }

The systems considered in Sections \ref{sec:subsol} and \ref{sec:modLP} 
are effectively one-dimensional, as all relevant quantities depend only on the coordinate $\rho$, with the exception of the metric factor $\tilde{\rho}$ (which may have an extra dependence on $v$ through $g(v)$, see \eqref{eq:rhot}).
These arise from a reduction to one dimension of the following four-dimensional Lagrangian\footnote{Note that we have rescaled the potential $V(\phi)$ in \eqref{eq:L4D} by a factor $2$ as compared with \eqref{lag3}.} \cite{Ferrara:1997tw,Goldstein:2005hq,Goldstein:2005rr,Goldstein:2014gta},
\begin{equation}\label{eq:L4D}
    L_{4D}= \frac12 \sqrt{-g}\left(R-2 G_{IJ}(\phi)\partial_\mu \phi^I \partial^\mu \phi^J-f_{ab}(\phi)F^{b}_{\mu\nu}F^{a\, \mu\nu}-\tilde{f}_{ab}(\phi)\tilde{F}^{b}_{\mu\nu}F^{a\, \mu\nu}-2V(\phi)\right)\,,
\end{equation}
corresponding to an Einstein-Maxwell theory with $N$ scalar fields $\phi^I$.
We take the target space metric $G_{IJ}$ to be constant and the four-dimensional metric to be of the form 
\begin{equation}\label{eq:le1d}
    ds_4^2=-a^2(r) dt^2+a^{-2}(r) dr^2+b^2(r) d\Omega_\alpha^2
\end{equation}
with 
\begin{equation}
    d\Omega_\alpha^2=d\theta^2+\ell_\alpha(\theta)^2 d\varphi^2\,, \quad \ell_\alpha(\theta)=\begin{cases} \,{\rm sin}(\theta)\,, \quad \alpha=1\,,\\
   \, 1\,, \quad \quad \quad  \alpha=0\,, \\
   \,{\rm sinh}(\theta)\,, \,\,\, \alpha=-1\,.
    \end{cases}
\end{equation}
We have adjusted the parameter $\alpha$ to make it coincide with the definition given in \eqref{eq:alpha}. The field equations read \cite{Goldstein:2014gta}, 
\begin{align}
    (a^2 b^2)^{\prime\prime}\,&=\,2(\alpha-2 b^2 V)\,\label{eq:un}, \\
    \frac{b^{\prime\prime}}{b}\,&=\, -(\phi^\prime)^2\,, \label{eq:dos}\\
    (a^2 b^2 G_{IJ}\phi^{J\prime})^\prime\,&=\,\frac12\left(\frac{\partial_I V_{\rm EM}}{b^2}+b^2 \partial_I V \right), \label{eq:tres}\\
    -\alpha +a^2(b^\prime)^2 +\frac12 (a^2)^\prime(b^2)^\prime& \,=\,-\frac{ V_{\rm EM}}{b^2}-b^2 V+b^2 a^2(\phi^\prime)^2 \,,\label{eq:cuatro}
\end{align}
with the last one being the Hamiltonian constraint. The previous equations are written in terms of $(\phi^\prime)^2=G_{IJ}\phi^{I\prime}\phi^{J\prime}$ and an electric-magnetic potential which is given by
\begin{equation}
    V_{\rm EM}(\phi)=f^{ab}(Q_{e,a}-\tilde{f}_{ac}Q^c_m)(Q_{e,b}-\tilde{f}_{bd}Q^d_m)+f_{ab}Q^a_m Q^b_m \,,
\end{equation}
where $Q_{e,a}$ and $Q_m^a$ are the electric and magnetic charges carried by $F^a$,
\begin{equation}
    F^a=f^{ab}(Q_{e,a}-\tilde{f}_{ac}Q^c_m) \frac{dt\wedge dr}{b^2} +Q^a_m \ell_{\alpha}(\theta) d\theta \wedge d\varphi \,.
    \label{Fa}
\end{equation}
A quick comparison of the line element \eqref{eq:le1d} with 
the line element \eqref{eq:lineelem2} (where we take $\lambda = 1$ and use ${\tilde \rho} (\rho, v) = h(\rho) g(v)$ with $g = \ell_\alpha $ and $dy = dt$),
\begin{equation}
   ds_4^2 =-\Delta dt^2+\Delta^{-1}(dr^2+h^2 d\Omega_\alpha^2) \,,
   \label{linOmal}
\end{equation}
gives 
\begin{equation}
    \Delta=a^2\,, \quad h=a \, b\,.
\end{equation}
The subsequent coordinate transformation $d r = e^{\Sigma (\rho)} \, d \rho, \, e^{\Sigma (\rho)} = h(\rho)$ brings the line element \eqref{linOmal} into the form \eqref{eq:lineelem2}.
Equations \eqref{eq:un} - \eqref{eq:cuatro} are consistent with \eqref{P1}, \eqref{h1} and \eqref{h22}, as follows. Recall that the equation for the field $A_\rho$ (eq. \eqref{P1}) supplies the equations of motion for the warp factor $\Delta$ as well as for the scalar fields $\phi^I$. Hence, \eqref{eq:tres} must come from the field equation for $A_\rho$. Similarly the $\Box\tilde{\rho}$ equation (eq. \eqref{h1}) corresponds to \eqref{eq:un}. The equation of motion
for $\Delta$ (which is part of \eqref{P1}) can be obtained from a combination of \eqref{eq:un}, \eqref{eq:dos} and  \eqref{eq:cuatro}. The consistency equation \eqref{h22} coincides with 
the Hamiltonian constraint \eqref{eq:cuatro}.
In what follows, 
we set
\bea
a^2=e^{2U} \quad , \quad b^2=e^{2\psi-2U} \;.
\eea
Then, inserting the line element \eqref{eq:le1d} and the expression for the field strength \eqref{Fa} into the Lagrangian \eqref{eq:L4D}
results in the following effective one-dimensional action \cite{Ferrara:1997tw}, 
\bea
   -S_{1D}&=&\int dr \left[ - \alpha + 
   e^{2\psi}\left((U^\prime)^2-(\psi^\prime)^2+ G_{IJ}\phi^{I\,\prime}\phi^{J\,\prime}+  V_{\rm EM}(\phi) \, 
   e^{2U-4\psi} \,+e^{-2U} V(\phi)\right) \right]\nonumber\\
   && +
   \int dr \left[ e^{2 \psi} \left( 2 \psi' - U' %
   \right) \right]'
   \,.
   \label{1Dact}
\eea
The total derivative term is irrelevant in constructing the equations of motion and ensuing dynamics, and we will ignore it henceforth.  
The 4D field equations can be derived from this effective 1D action as follows. Choosing the 1D field degrees of freedom to be $e^{\psi + U}$, $e^{\psi-U}$ and ${\phi^I}$, we see that the equations \eqref{eq:dos} and \eqref{eq:tres} are obtained by the Euler-Lagrange variation of the action w.r.t $a^2 b = e^{\psi+U}$ and $\phi^I$ respectively. Given that the 1D Lagrangian does not explicitly depend on `time'\footnote{Here we take the radial co-ordinate to be `time'.}, its Hamiltonian 
$\mathcal{H}$ is constant on-shell. Hence the fourth 4D field equation \eqref{eq:cuatro} is recognizable as the Hamiltonian constraint $\mathcal{H}=0$. The first field equation $\eqref{eq:un}$ is obtained by substituting this constraint into the expression obtained by the Euler-Lagrange variation w.r.t $b= e^{\psi-U}$.  
Thus  we can regard  the solution space of the 1D action as being generated by field configurations satisfying \eqref{eq:un}, \eqref{eq:dos} and \eqref{eq:tres}, which we will refer to as the equations of motion (EOMs). 
Consequently, using a relabeling of the field degrees $A= a^2 b$ and $B=b$, we rewrite the bulk 1D action  which generates the EOMs 
when supplemented by the Hamiltonian constancy $\mathcal {H} =0 $
 as  
  \begin{eqnarray}\label{1Dact1}
-S_{1D}=\int d r
   \left(-A'\, B'\,+ A\, B\, G_{IJ} \phi^{I\,\prime}\phi^{J\,\prime}+  \frac{V_{\rm EM}(\phi)}{B^2} + B^2 V(\phi)\right) 
   \,.
\end{eqnarray}
In \eqref{1Dact1}, we have dropped the constant $\alpha$ as it does not affect any of the equations of motion.

For the purpose of studying 1D systems represented by actions of the form above, we do not impose the Hamiltonian constraint necessary for embedding the 4D solution space in the 1D solution space.  Instead, given that the 1D action is expressed by a time-independent Lagrangian, the Hamiltonian on an on-shell trajectory is a constant of motion, equal to say $ \beta$:
\begin{align}
    \beta - \alpha +a^2(b^\prime)^2 +\frac12 (a^2)^\prime(b^2)^\prime& \,=\,-\frac{ V_{\rm EM}}{b^2}-b^2 V+b^2 a^2(\phi^\prime)^2 \,.\label{eq:cuatrogeneral}
\end{align}
This  leaves  \eqref{eq:dos} and \eqref{eq:tres} unchanged, but modifies \eqref{eq:un} to
\bea
(a^2 b^2)^{\prime\prime}\,&=\,2(\alpha - \beta -2 b^2 V)\,.
\label{neweom1}
\eea
The Hamiltonian constraint $\mathcal{H}=0$ is simply enforced by $\beta = 0$ to recover  \eqref{eq:un}.

A classical system  represented by \eqref{1Dact1} is defined to be Liouville integrable if one can identify $n$ independent\footnote{Independence here implies that given $n$ integrals of motion $\{P_i | 1 \leq i \leq n\}$, $ dP_1 \wedge dP_2 \wedge \dots \wedge dP_n \neq 0$. Thus functions of the phase space variables that are constant throughout the phase space 
are not counted as integrals of motion for our purpose.} commuting first integrals of motion\footnote{A function of phase space variables is a first integral of motion iff it is conserved on each on-shell trajectory and its value is independent of the choice of initial conditions. This distinguishes it from a constant of integration obtained by solving an EOM as said constant depends explicitly on the choice of the initial time.} (i.e. conserved currents possessing vanishing Poisson brackets with each other) on its phase space, where $n$ is the dimension of the configuration space. One of the strategies for identifying these integrals is to write down a Lax pair with matrices $L$ and $M$ which satisfy
\begin{align}
    \frac{dL}{dt}=[L,M]
    \label{Laxcond}
\end{align}
when evaluated on-shell, and then extract $n$ independent integrals from $L$ which have to be shown to be in involution, i.e. Poisson commute with each other.  Hence, the existence of a Lax pair is only a necessary condition for integrability. For the remainder of the text, we will often work with a specific `canonical' construction of the Lax pair based on the following argument:
if we can identify $n$ 
independent commuting integrals, $\{P_i | 1\leq i \leq n \}$, one of which is the 
Hamiltonian, 
one can readily write down an $n\times n$ matrix Lax pair
\begin{eqnarray}\label{laxaction}
L_{ij}&=& P_i \delta_{ij}\,, \nonumber\\
M&=& I_{n\times n} \,,
\end{eqnarray}
which clearly satisfies the Lax condition \eqref{Laxcond}.
Hence, the existence of this canonical Lax pair is both a necessary and sufficient condition to infer the integrability of the system.

In the absence of a potential, $V(\phi)=0$, we see that the EOM \eqref{neweom1}, 
\beq
(AB)''=2 \left( \alpha - \beta \right) \;,
\eeq 
simply eliminates $A$ in terms of $B$ as 
\beq \label{c1}
A = \frac{(\alpha-\beta) r^2 + c_1 r + c_2}{B} \;,
\eeq 
where $c_1$ and $c_2$ are the constants of integration of the EOM. The constant $c_2$ can be eliminated by choosing the origin of the radial coordinate,  which is a gauge choice leaving us with the  conserved charge $c_1$ characterising on-shell orbits in phase space.   Hence, $A$ is completely determined in terms of $c_1$ and $B$, and the remaining degrees of freedom correspond to the $\phi^I$. 
Now each scalar degree of freedom $\phi^I$ can be associated with a constant of motion, 
\bea
\Pi_I = P_I -  \int_{+\infty}^{r} \frac{\partial_I V_{\rm EM}(\phi)} {B^2} d \tilde{r},\,
\eea
where $P_I =  2 \,A\,B\, G_{IJ}(\phi ^J)'$ is the momentum associated with the scalar degree of freedom $\phi^I$.  
This is evident from 
the EOM \eqref{eq:tres} guaranteeing $d\Pi_I / dr = 0$, and further,  it is easy to observe that the Poisson bracket vanishes, $\{\Pi_I, \Pi_J\}=0$, thus
 verifying that these canonical momenta constitute a set of independent commuting conserved functions of phase space variables. These functions can only be shown to be first integrals of motion iff we can demonstrate that they take a finite on-shell value and do not take the same value for all solutions in the $\mathcal{H}=0$ subclass.  In order to make this argument, we recall that each such solution, representing a 4D gravitational configuration, is associated with a fixed asymptotic limit for its fields  
  as $r \rightarrow + \infty$.  We evaluate $\Pi_I$ to establish its finiteness in order to determine its viability as a first integral.  As an illustrative example, for $V(\phi)=0$, one possible set of 4D asymptotic conditions corresponds to flat Minkowski space-time which translates to the asymptotic limits  $A, B \approx r $,  while $\phi^I \approx \phi^I_{\infty} + \frac{\Sigma^I}{r}+\dots$. Using $\frac{d\Pi_I}{dr}=0$, we see that \beq \Pi_I = P_I¦_{\infty}= - 2 G_{IJ}\Sigma^J .\eeq Therefore, in this case $\Pi_I$ assumes an independent finite on-shell value. This check for finiteness must be passed on every solution in the $\mathcal{H}=0$ class to establish it as a first integral of motion.  Note that our restriction of the 1D solution space to a subspace upliftable to 4D implies $\mathcal{H}= \beta =0$, which lowers the number of independent degrees of freedom by one. Hence, given a successful finiteness check, $c_1$ and the $\Pi_I$ constitute a set of independent commuting integrals of motion, whose cardinality is equal to the number of degrees of freedom, which verifies the integrability of the system.\footnote{Invoking the Hamiltonian constraint $\mathcal{H} = \beta = 0$ for the 1D solution space to be lifted to a 4D one, has the consequence of $\mathcal{H}$ being zero all over the 4D solution subspace and hence it is no longer 
 regarded as an integral of motion.}

A similar argument holds in the case $V(\phi)\neq 0$, where one can write down a set of constants of motion as 
\beq 
\Pi_I = P_I -  \int_{+ \infty}^r \left(\frac{\partial_I V_{\rm EM}(\phi)} {B^2} +  B^2 \partial_I V(\phi)\right) d \tilde{r} \;.
\label{pii}
\eeq 
However, in this case, the EOM 
\beq \label{intecon}
(A B)'' = 2 \left( \alpha - \beta - 2 B^2 V(\phi) \right)
\eeq
is not trivially solvable in contrast to the $V(\phi)=0$ case and may not necessarily yield a conserved 
charge.
Hence, the degrees of freedom are generically $A, B$ and the scalar fields $\phi^I$. 
In this case, we have a variety of possible asymptotic fall-offs which may result in the $\Pi_I$ being divergent asymptotically, in which case we need to extract regularised values for them, if they are to be counted as first integrals.  As an example, say we consider fall-offs defined by 
 $A B\approx r^m$ and $\left( \phi^I \right)' \approx \frac{\Sigma^I}{r^l}+ \dots$.
  Hence, we can evaluate the $\Pi_I$ to be
 \beq
\Pi_I = P_I¦_{\infty}= 2  \, r^{m-l}G_{IJ}\Sigma^J.
\eeq
For $m-l > 0$
the $\Pi_I$ values diverge asymptotically\footnote{This is par for the course for numerous physical quantities such as the stress tensor in the presence of a cosmological constant, which must then be regularised.} and we regularise these values by peeling of the divergent $r^{m-l}$ power to obtain
\bea
\Pi_{I\, {\rm reg}}= 2 \, \, G_{IJ}\Sigma^J \,.
\label{PiregV}
\eea 
 As in the $V(\phi)=0$, the finiteness check over each solution determines whether  the $\Pi_I$ qualify as independent integrals of motion, which Poisson commute with each other.  

Here again,  as in the vanishing $V(\phi)=0$ case,  restricting to the subspace of the 1D solution space that is upliftable to the solution space of the parent 4D Lagrangian, requires us to impose the null Hamiltonian constraint \eqref{eq:cuatro}, which reduces the number of degrees of freedom by one.  Hence the cardinality of the set of integrals constituted of the $\Pi_I$ (if they qualify as first integrals) is one less than the number of the degrees of freedom. The latter is the required number of integrals of motion for establishing Liouville integrability.  The presence of an extra integral of motion has to be evaluated on a case-by-case basis. 
Below, we will illustrate our conclusions above with specific 1D examples 
by using both analytic and 
machine learning (ML) techniques to construct explicit Lax pairs and conserved currents for these systems.

\section{Lax pairs for 1D systems \label{sec:scla1D}}

In the following, we consider systems described by the one-dimensional action \eqref{1Dact}.
For the remainder of our discussion we will focus only on the bulk term, as the total derivative boundary term does not affect system dynamics. For the same reason, we will also drop the term proportional to $\alpha$.
For simplicity, we restrict the discussion to the case of one Maxwell field
sourced by an electric charge $Q_0$ only, in which case $V_{\rm EM}(\phi) =
\frac14 f(\phi) Q_0^2$.
We then consider specific choices of the coupling $f$ and of the potential $V$, and we
construct Lax pair matrices $(L,M)$ for these systems. Systems described by the one-dimensional action \eqref{1Dact} with $\alpha =0$ were considered in the context of Nernst branes in \cite{Barisch:2011ui}.

Note that a system may possess a description in terms of several Lax pairs.
In Appendix \ref{sec:A} we exhibit two families of Lax pairs for 1D systems described by a Hamiltonian of the form $2 \mathcal{H}(p,q) = p^2 + 2 V(q)$.

\subsection{No scalar field}\label{sec:nosc}

Let us first consider the case when there is no scalar field present in \eqref{1Dact} or equivalently, \eqref{1Dact1}. 
The 1D action in this case represents 2 degrees of freedom. The Hamiltonian $\mathcal H$ in this case is clearly independent of $A$, rendering $P_A$ to be a constant of motion. Hence, we have two mutually commuting independent conserved currents, $\mathcal H$ and $P_A$, corresponding to the 2 degrees of freedom, thus verifying Liouville integrability. Alternatively, if we restrict to solutions that are upliftable to 4D gravitational configurations, we must fix $\mathcal{H}=0$.  This reduces the number of degrees of freedom to one and $P_A$, being conserved, is enough to guarantee integrability. We will illustrate this in the example below.

Here, we choose the coupling $f$ and the potential $V$ in \eqref{1Dact} to be constant.
Setting $V=-\frac34 h_0^2$ and $f=1$, 
we obtain 
\begin{equation}
\label{ac1Dqh}
   -S_{1D}=\int dr \, e^{2\psi}\left[(U^\prime)^2-(\psi^\prime)^2+\frac14 Q_0^2 e^{2U-4\psi}-\frac34 h_0^2 e^{-2U}\right] \,.
   \end{equation}
This system, which depends on the two parameters $(Q_0, h_0)$, admits a Lax pair $(L,M)$ that satisfies \eqref{Laxcond} when evaluated on-shell. Note that the $Q_0=0$ model admits a solution that can be lifted up to the $AdS_4$-Schwarzschild solution discussed in Section \ref{sec:schwarzads}, after imposing \eqref{eq:cuatrogeneral} with $\alpha =1$ and $\beta=0$.

Setting $A=e^{\psi+U}, B= e^{\psi-U}$, the action \eqref{ac1Dqh} becomes 
\begin{equation}\label{ac1Dqh2}
   -S_{1D}=\int dr \left[-A^\prime B^\prime+\frac{Q_0^2}{4B^2} -\frac34 h_0^2 B^2\right]\,,
\end{equation}
with the Hamiltonian being 
\bea
{\cal H} = -P_A P_B
- \frac{Q_0^2}{4B^2}+ \frac34 h_0^2 B^2\;.
\label{cal H}
\eea
A canonical Lax pair for this system is given in terms of the integrals of motion $P_A = - B^\prime$ and $\mathcal{H}$ as
\bea
L = \begin{pmatrix}
P_A & 0 \\
0 & \mathcal{H} 
\end{pmatrix} \;\;\;,\;\;\; M = \mathbb{I}_2 \;.
\eea
An alternative Lax pair is given by 
$3 \times 3 $ matrices of the form 
\begin{equation}\label{ourLax}
    L=\begin{pmatrix} 0 & -A^\prime & G_2 \\
    B^\prime  & 0 & 0\\
    G_1 & 0 & 0
    \end{pmatrix}\,, \quad M=\begin{pmatrix} 0 & 0 & 0\\
    0 & 0 & 0\\
    0 & -\frac{G_1^\prime G_2+G_2^\prime G_1}{B^\prime G_2} & \frac{G_2^\prime}{G_2}  
    \end{pmatrix}
\end{equation}
with 
\begin{equation}
    G_1=-\frac{Q_0}{2B}+\frac{\sqrt{3}}{2}h_0 B\,, \quad G_2=\frac{Q_0}{2B}+\frac{\sqrt{3}}{2}h_0 B\,,
\end{equation}
where $L$ satisfies ${\rm Tr}(L^2)=2{\cal H}$. 

Note that the action in this case has a translation symmetry, $A\rightarrow A+ c$, where $c$ is an arbitrary constant. This implies that the 2 conserved currents making up the canonical Lax pair define each on-shell trajectory uniquely modulo this symmetry. For the family of $AdS_4$ Schwarschild black holes with $m$ representing the mass, for which $A = r - 2 m + \frac{r^3}{L^2}$ and $B=r$ and hence $P_A = -1$, the translation symmetry in $A$ generates all such black hole solutions from a pure $AdS_4$ solution corresponding to $m=0$. Hence, the mass information is not present in the value of $P_A$.  In the presence of scalars, the action \eqref{1Dact1} does not admit this translation symmetry and hence, the constant $c_1$ in \eqref{c1} which plays the role of the mass cannot be eliminated by symmetry and therefore is a constant of motion that characterises orbits.  

\subsection{One scalar field}\label{sec:onescalarfield}

Next, let us consider the case when there is one scalar field $\phi$ present in \eqref{1Dact}, with the coupling $f$ and the potential $V$ given by
\bea
 f(\phi) &=& 2 e^{2 \phi} \,, \nonumber\\
 V(\phi) &=& - \left(h_0 h_1 + \frac12 \left(h_0 e^\phi+h_1 e^{-\phi}\right)^2 \right) \,,
\eea
where $h_0, h_1$ denote constants. This choice of $f$ and $V$ is the one that occurs in the model based on $F (X) = - i X^0 X^1 $ discussed in \cite{Barisch:2011ui}. The resulting 1D action reads
\bea\label{eq:eqonef}
   -S_{1D}&=&\int dr e^{2\psi}\left[(U^\prime)^2+(\phi^\prime)^2-(\psi^\prime)^2+\frac12 e^{2U+2\phi-4\psi} Q_0^2 \right. \nonumber\\
   && \left. \qquad \qquad 
- e^{-2U} \left(h_0 h_1 + \frac12 \left(h_0 e^\phi+h_1 e^{-\phi}\right)^2 \right)\right]\,. 
\eea
Labelling 
\begin{eqnarray}\label{basisc}
 \log A &=& U +\phi + \psi \,, \nonumber \\
    \log B &=& -U -\phi + \psi \,, \nonumber \\
     C &=& - U + \phi + \psi \,,
\end{eqnarray}
we can rewrite the action as
\begin{eqnarray}\label{abcQh0h1}
   -S_{1D}&=&\int dr \left[- \frac12 A' B' - \frac12 (A' B + A B') C' + \frac12 A B (C')^2 +\frac12 \frac{Q_0^2}{B^2}\right.\nonumber \\ 
&& \qquad \qquad  \left. - \left(2  h_0 h_1 B e^C +\frac12 h_0^2 e^{2C} + \frac12 h_1^2 B^2\right)\right]\,.
\end{eqnarray}
This system depends on the 3 parameters $(Q_0, h_0, h_1)$.
The canonical momenta are given by
\bea 
P_A + \frac{B'+ B\, C'}{2}&=&0,\nonumber\\
P_B + \frac{A'+ A\, C'}{2}&=&0,\nonumber\\
P_C + \frac{(A\,B)'}{2}- A\, B\, C'&=&0 .
\label{pabcexp}
\eea 
The Hamiltonian, which is conserved on-shell, takes the form
\begin{eqnarray}
\mathcal{H} =&- \frac32 P_A P_B +\frac{AP_A(A P_A-2P_C)+BP_B(B P_B-2P_C)+P_C^2}{4AB}\nonumber\\
&-\frac{Q_0^2}{2B^2} + \left(2  h_0 h_1 B e^C +\frac12 h_0^2 e^{2C} + \frac12 h_1^2 B^2\right) \;.    
\end{eqnarray}
Hamilton's equation yield
\bea
\label{Pabcpr}
P'_A + \frac{B' C' - B\,C'^2}{2}&=&0,\nonumber\\
P'_B +2  h_0 h_1 e^C+\frac{Q_0^2}{B^3}+ h_1^2 B + \frac{A' C' - A\,C'^2}{2}&=&0,\nonumber\\
P'_C +  h_0^2 e^{2C} + 2  h_0 h_1 B e^C&=&0.
\eea

Using \eqref{pabcexp}, one readily verifies that the following combination is a on-shell conserved quantity,
\bea\label{eqccxx2}
  J = A(r) B(r) + \int_{+\infty}^r  \left[ A P_A + B P_B + P_C \right] \, d {\tilde r}\;,
  \eea
The value of $J$ is then given by 
\bea\label{eqccxx3}
  J = A(+\infty) B(+\infty) .
  \eea
Noticing that $A B = a^2 b^2$ represents metric factors in the four-dimensional line element \eqref{eq:le1d},
we immediately see that the regularised value of $J$ will be determined purely by the asymptotic metric behaviour and fall-off, and varies for solutions with different asymptotia in the $\mathcal{H}=0$.  Consequently, it qualifies as an integral of motion for all such solutions. It does not, however, commute with the 
current $\Pi_{\rm reg}$ given in \eqref{PiregV}.
In terms of the fields in \eqref{basisc}, $\Pi$ is expressed as
\bea
\Pi = P_C + A \, P_A - B \, P_B - \int_{+\infty}^r  \left[  \frac{Q_0^2}{B^2} - \left( h_0^2 \, e^{2C}- h_1^2 \, B^2 \right) \right] \, d {\tilde r}\;.
\label{picurabc}
\eea
  We have not been able to construct an additional independent conserved current that Poisson commutes with $\Pi$ or $J$ and hence cannot verify the Liouville integrability of the system in the general case when the parameters 
$(Q_0, h_0, h_1)$ are all non-vanishing.

In the specific case when $h_0=0$, it is easily seen from \eqref{Pabcpr} that the momentum $P_C$ becomes a 
  constant of motion. Moreover,  
  $\{P_C, \Pi\} = 0$.  If for each solution in the $\mathcal{H}=0$  class, $P_C$ and $\Pi$ take finite on-shell values (possibly after appropriate regularisation) which are not constant all over the $\mathcal{H} =0$ solution class, they qualify as conserved currents. From \eqref{picurabc}, we see that the difference between their values depends purely on the metric fall-offs and generically they are independent. 
  Therefore, given successful finiteness checks for the $P_C$ and $\Pi$ over the entire $\mathcal{H}=0$ class, we have three independent commuting integrals of motion, verifying Liouville integrability.

   This analytic construction of currents is in accord with our machine learning experiments to which we turn in the next section.

\section{Machine learning integrability}\label{sec:nums}

In this section we use machine learning (ML) techniques to search for Lax pairs in the 1D setup. 
A simple way to go about our ML implementation is to frame the search for these quantities as a neural network (NN) optimisation problem.
 ML searches of Lax pairs and $r$-matrices in classical and quantum integrability have been initiated in \cite{Ishikawa_2021,Krippendorf:2021lee,daigavane2022learning,Lal:2023dkj}. The efficacy of problem solving depends on the neural network architecture parameters (number of layers, neurons per layer, activation function, optimiser, etc). The mathematical principle underlying this approximation technique is the universal approximation theorem \cite{FUNAHASHI1989183,Cybenko1989,HORNIK1989359}, which guarantees that a continuous function on a compact subset of $\mathbb{R}^n$  can be arbitrarily well approximated by a neural network in the limit of  infinite NN length or width.  Therefore, in practice, the accuracy of the approximation is bounded 
 by the finite extent of the NN. And so, a critical component of this approach consists in setting up a NN that upon training, produces a certain desired accuracy on the numerical Lax pair approximation. This accuracy is a value that we deem acceptable for deciding whether the system under consideration could be integrable or not. 
In this section we implement an ML search for both canonical and non-canonical Lax pairs.

In our particular instance, we adopt an unsupervised learning approach wherein we consider a neural network that takes points in phase space as inputs and produces the entries for the matrices $L$ and $M$ as outputs. We  do not know $L$ and $M$ a priori, and hence start with a randomly initialised neural network, i.e. defined with randomly fixed initial parameters, and subsequently update the NN parameters\footnote{We collectively refer to all the NN network parameters heuristically by a single variable $\theta$ for notational simplicity.} $\theta$ to attain values optimised to minimise deviation from the Lax pair condition (or the conservation law for a given numerical current). The quantity that will be minimised is known as the loss function $\mathcal{L}$
and the optimisation procedure follows the gradient descent method given by
\begin{equation}
    \theta_{i+1}=\theta_i - \ell_r \nabla_\theta \mathcal{L}(X,X^\prime,\theta)|_{\theta=\theta_i}.
\end{equation}
Here $\ell_r$ is called the learning rate\footnote{The learning rate is part of the model design and similar to the neural network architecture it is chosen by hand. Quantities of this type, that are not optimised during the training process, are referred to as hyperparameters.} and measures the rate of convergence or the convergence step size of $\theta$ towards optimal values. 
The variables $X$ and $X^\prime$ respectively denote configuration space variables and their corresponding time\footnote{Here `time' simply refers to the independent variable w.r.t which the derivatives of configuration space variables are taken to define velocities.} derivatives, which together constitute the NN input.
For a 1D system with $k$ degrees of freedom (a $k$-dimensional configuration space) described by 
\begin{eqnarray}    
X&=&(\phi_1,\phi_1^\prime,\phi_2,\phi_2^\prime,\ldots,\phi_k,\phi_k^\prime),\nonumber\\
X^\prime&=&(\phi_1^\prime,\phi_1^{\prime\prime},\phi_2^\prime,\phi_2^{\prime\prime},\ldots,\phi_k^\prime,\phi_k^{\prime\prime}),
\label{XXp}
\end{eqnarray}
data is sampled randomly from within a cube $\phi_i,\phi_i^\prime\in\,[-1,1]$. We take 100.000 points for each of our experiments and split the data into $80\%/20\%$ subsets for training and validation. 
At this point we mention that for all ML experiments in this paper, the phase space point sampling was performed in \texttt{Mathematica} with the ML implementations done in \texttt{Tensorflow} \cite{tensorflow2015-whitepaper}.

As mentioned before, minimisation of the loss function must correspond to finding a pair of matrices $L$ and $M$ corresponding to minimum deviation from the Lax pair equation \eqref{Laxcond}. This motivates a choice of the loss function to be 
\begin{equation}
    \mathcal{L}_{\rm Lax,1}=\sum_{ij}|(dL_{ij}/dt)-[L,M]_{ij}|^2\,,
\end{equation}
or  equivalently, 
\begin{equation}
    \mathcal{L}_{\rm Lax,2}=\sum_{ij}\left|\frac{{\rm Exp}(dL_{ij}/dt)}{{\rm Exp}([L,M]_{ij})}-1\right|^2
\;.
\end{equation}
The values of $dL_{ij}/dt$ at each point $X$ are computed via the chain rule 
\begin{equation}
\frac{dL}{dt}=\frac{dL}{d\phi}\phi^\prime + \frac{dL}{d\phi^\prime}\phi^{\prime\prime}\,,
\end{equation}
where $dL/d\phi$ and $dL/d\phi^\prime$ are derivatives of NN outputs with respect to inputs, and are obtained via automatic differentiation, while the values of $\phi^{\prime\prime}$ are supplemented from the equations of motion. 

There are two principal caveats to this approach. First, the above extremisation could produce trivial solutions such as $L_{ij}=0$. This can be avoided in one of two ways. The simplest way is to add a regulator to the loss function. 
A systematic way to do this comes from 
observing that in a 1D integrable system with $n$ degrees of freedom, an $n \times n$ Lax pair produces an infinite set of conserved quantities, $\{F_k = {\rm Tr} L^k,\, k \in \mathbb{Z}$\}, of which at most only $n$ are independent and Poisson commuting, i.e. have a vanishing mutual Poisson bracket.  Requiring the Hamiltonian to be one of these $n$ commuting integrals, eliminates trivial solutions. In our setup, we accordingly choose the Hamiltonian to be the trace of the square of the $L$-matrix, and implement this by introducing the following Hamiltonian loss constraint,
\begin{equation}
\mathcal{L}_{\cal H}=|{\rm Tr}(L^2)- {\cal H}|^2,
\end{equation}
so that the total loss function is the weighted linear combination,
\begin{equation}
    \mathcal{L}=\alpha_{\rm Lax} \mathcal{L}_{\rm Lax}+\alpha_{\cal H} \mathcal{L}_{\cal H} ,
    \label{eq:Loss}
\end{equation}
with $\alpha_{\rm Lax}$ and $\alpha_{\cal H}$ being positive coefficients, which we choose to be equal to 1 for all experiments conducted in this section. The second caveat is that for a given matrix dimensionality,  the afore defined loss function minimization does not uniquely fix the Lax pair.  This is because, given a Lax pair, one can generate an equivalent Lax pair satisfying the Hamiltonian loss constraint by a rescaling $\lambda \in \mathbb{R}$ and a $GL(n,\mathbb{R})$ transformation $U$,
\begin{equation}
    L \rightarrow \lambda \,U L U^{-1}\,, \quad M\rightarrow U M U^{-1}-\frac{dU}{dt}U^{-1} \;.
\end{equation}
 One could argue that this equivalence is not necessarily an obstacle for our searching technique, as we are only interested in proving the existence of a Lax pair and this caveat only assumes significance if we want to search for a specific 
Lax pair. 
Further, as explained above, the results of our search algorithm are never exact, and so one must have a consistent numerical threshold tolerance, within which the extremum value of the loss function can be taken to be zero, for all intents and purposes, and the existence of this extremum can be taken as a convincing indicator of integrability. 

In order to construct numerical Lax pair approximators, we start by considering the 1D Hamiltonian given in \eqref{eq:hamiltonian1D}. For concreteness we contemplate three different potentials for our experiments: the Liouville potential $V(q)=g^2 e^{-q}$, the de Alfaro-Fubini-Furlan (DFF) potential $V(q)=\frac12 \omega^2 q^2 +g^2/q$ as well as a linear potential $V(q)=g^2 q$. Lax pair candidates for these systems are known and are discussed in Appendix \ref{sec:A}. 

For all the ML experiments performed in this section, we have employed a deep neural network architecture with three hidden layers, each containing 350 neurons with a hyperbolic tangent activation function. 
This NN is to produce a Lax pair $(L,M)$ as output.  Schematically, the training protocol is presented in Figure \ref{fig:arch}.

\begin{figure}[h!]
\centering
\tikzstyle{sensor}=[draw, fill=blue!20, text width=5em, 
   text centered, minimum height=2.5em]
\tikzstyle{sensor2}=[draw, fill=green!20, text width=5em, 
    text centered, minimum height=2.5em]
\tikzstyle{ann} = [above, text width=5em, text centered]
\tikzstyle{wa} = [sensor, text width=9em, fill=red!20, 
   minimum height=6em, rounded corners]
\tikzstyle{sc} = [sensor, text width=13em, fill=red!20, 
   minimum height=10em, rounded corners]
\begin{tikzpicture}
   \node (wa) [wa]  {NN};
   \path (wa.west)+(-3.2,0) node (asr1) [sensor] {$X$};
   \path (wa.west)+(-3.2,-2) node (asr2) [sensor] {$X^\prime$};
      
    \path (wa.south)+(0,-1.5) node (vote0) [sensor2] {$\mathcal{L}$};
    
   \path (wa.west)+(6.0,-1.0) node (vote1) [wa] {$L,\,M$};

    \path [draw, ->] (asr1.east) -- node [above] {} 
       (wa.180) ;
   \path [draw, ->] (asr2.east) -- node [above] {} 
        (vote0.180);
    \path [draw, ->] (vote0.north) -- node [above] {} 
        (wa.south);
     \path [draw, ->] (wa.east) -- node [above] {} 
        (vote1.160);
    \path [draw, ->] (vote1.180) -- node [above] {} 
        (vote0.east);
\end{tikzpicture}
\caption{\emph{Flow chart of the Lax pair training process.
The neural network takes points in phase space as its inputs, and in training,  minimises the loss function \eqref{eq:Loss}, using the EOMs, to output
a numerical approximation for a Lax pair $L$, $M$ at each point.
}}
   \label{fig:arch}
\end{figure}

\begin{figure}[h!]
\centering
\includegraphics[scale=.37]{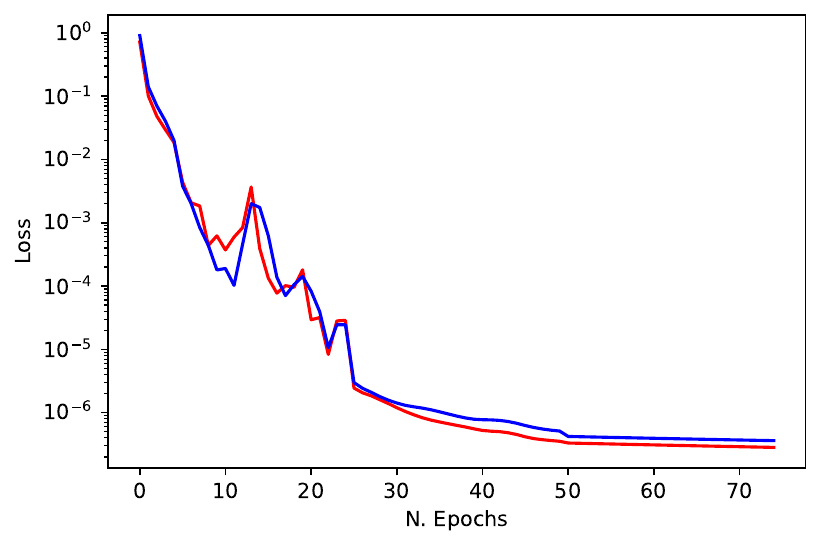}
 \includegraphics[scale=.37]{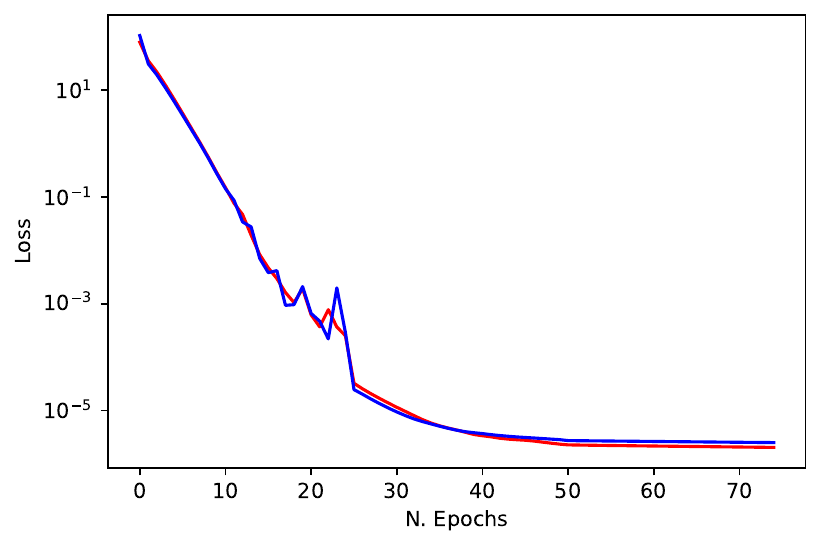}
  \includegraphics[scale=.37]{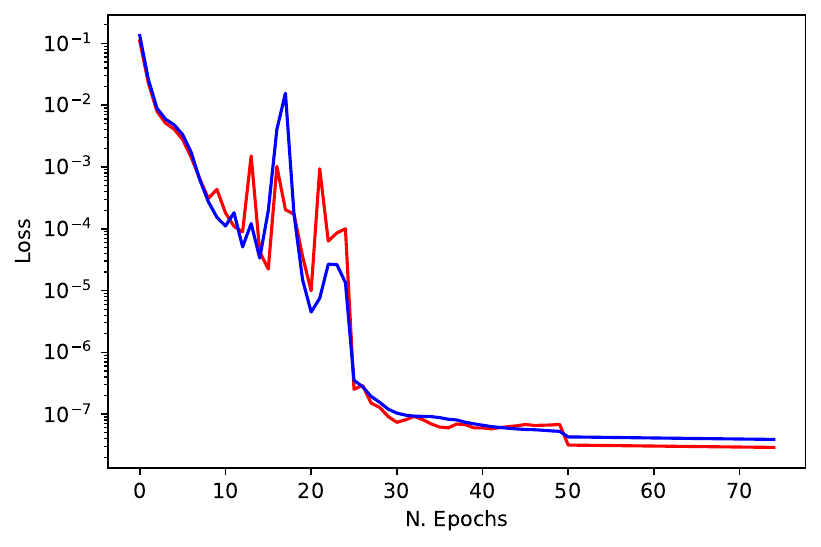}
\caption{\small\textit{Training process for three different potentials. Left: Liouville potential $V(q)=e^{-q}$. Center: de Alfaro-Fubini-Furlan (DFF) potential $V(q)=\frac12 q^2 +1/q$. Right: A linear potential $V(q)=q$}.}
  \label{fig:1DHams}
  \vspace{-10pt}
\end{figure}

Throughout the training process we implement the following learning rate schedule \cite{lrpaper}. We start training with a learning rate $\ell_r$ of $10^{-3}$ which we keep constant during the first 25 epochs of training. Afterwards, the $\ell_r$ is updated to $10^{-5}$ for the next 25 epochs and finally set to $10^{-6}$ for the remainder of the training process. Sampling of phase space elements is done by setting $\omega$ and $g$ to 1. After the training process, where we minimised the loss function \eqref{eq:Loss}, we obtain losses of around $10^{-6}$, which correspond roughly to deviations of $10^{-3}$ away from zero for the expression $dL/dt-[L,M]$. This is shown in Figure \ref{fig:1DHams}.

\begin{figure}
\centering
\includegraphics[scale=.37]{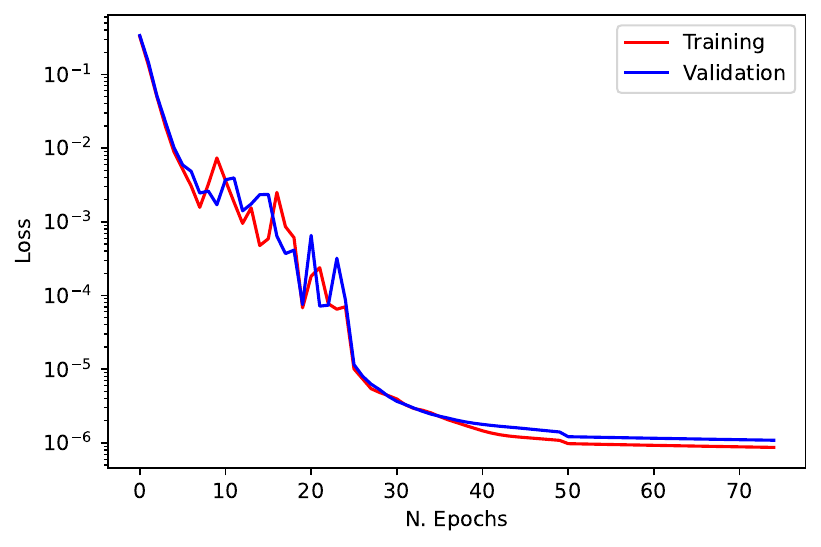}
 \includegraphics[scale=.37]{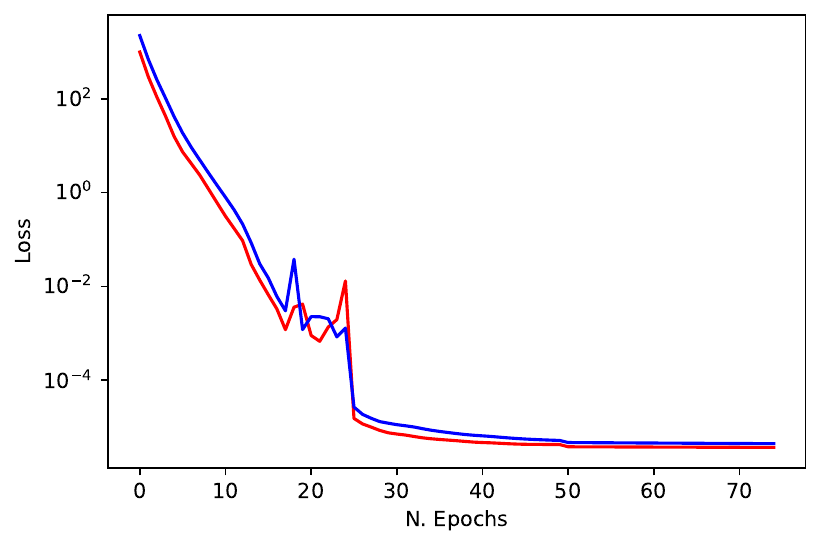}
  \includegraphics[scale=.37]{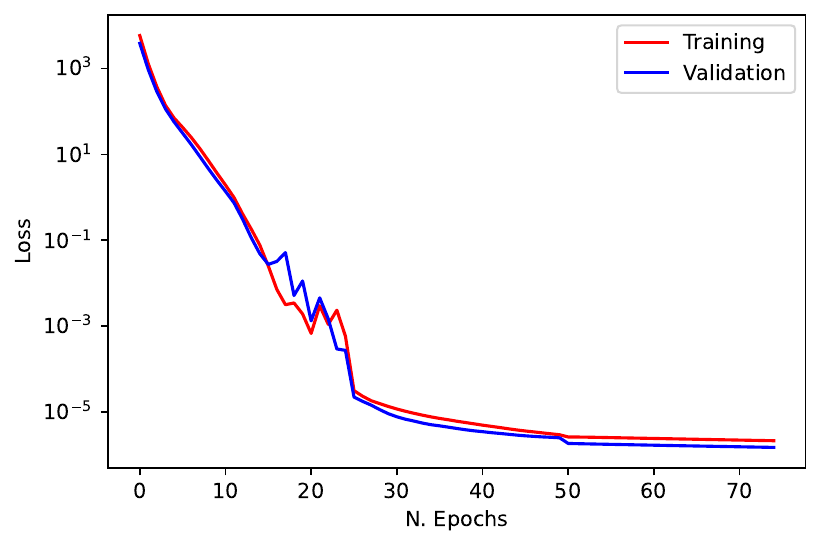}
\caption{\small\textit{Numerical Lax pair training in the no scalar field model described by 
\eqref{eq:rescaled}.
Left: Case (a) $h_0=0$, $Q_0=1$. Center: Case (b) $h_0=1$, $Q_0=0$. Right: Case (c) $h_0=Q_0=1$.
}}
  \label{fig:1DModels}
\end{figure}

\vspace{30pt}
Having obtained numerical Lax pairs for the models with a single degree of freedom we now proceed to explore the no scalar field model. Such a model was introduced in Section \ref{sec:nosc}. In Appendix \ref{sec:A} we have obtained a $4 \times 4$ analytic Lax pair for a different no scalar model \eqref{eq:rescaled} in the limiting case corresponding to $h_0=0$. For the purpose of studying the classical integrability of the 1D solution subspace that is upliftable to a 4D gravity by imposing $\mathcal{H}=0$, both these models are physically equivalent.  We now implement an ML search for $4 \times 4$ Lax matrices, and perform the training along similar lines to the methodology described above. We have considered three cases, namely, (a) $h_0=0, Q_0=1$, (b) $h_0=1, Q_0=0$ and (c) $h_0=Q_0=1$. The results are shown in Figure \ref{fig:1DModels}. 

{From} \eqref{eq:rescaled} we see that in case (a) we have a free particle described by $\psi$, while $U$ is governed by a Liouville type Hamiltonian with $h_0=0$.   After training we obtain losses of the order of $10^{-6}$. In cases (b) and (c), where $h_0\neq 0$, we observe that the network training begins at much higher losses of order $10^3$ and $10^4$ respectively.  However, after training, the losses are of the order of $10^{-6}$, the same order of magnitude as in the case (a). This suggests that cases (b) and (c) are as likely to be integrable as (a). Viewing the $h_0$ term as a deformation of the integrable system in case (a), this is in accord with the observation of \cite{Krippendorf:2021lee,Lal:2023dkj} which states that if the post training losses for an integrable system and its deformation are of the same order of magnitude, then the deformation has a good chance of being integrable as well. This is indeed borne out by the analytic results of Section \ref{sec:nosc}. 

\begin{figure}[h!]
\centering
\includegraphics[scale=.37]{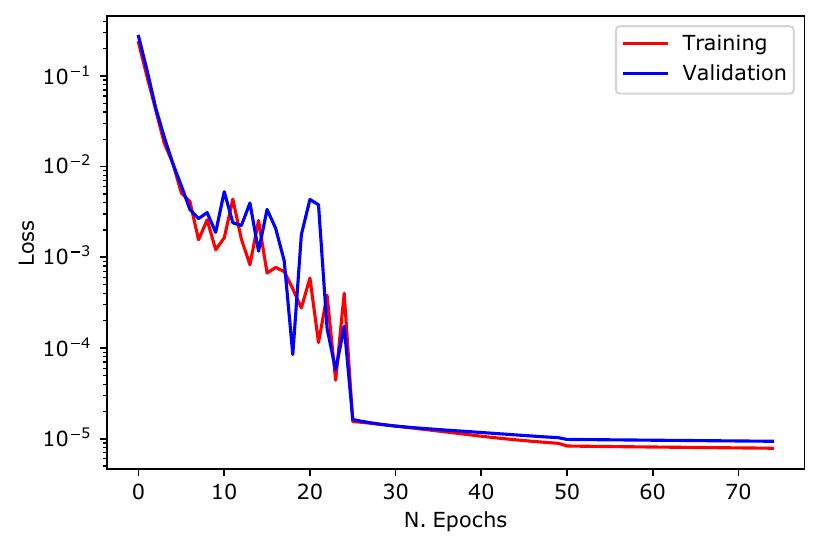}
 \includegraphics[scale=.37]{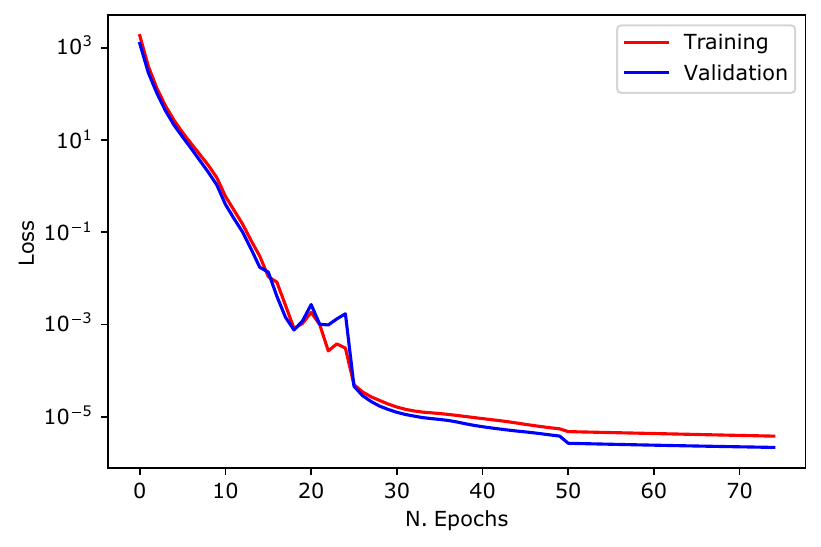}
\caption{\small\textit{Numerical Lax pair training in the one scalar field model (see 
\eqref{actuppsi}).
Left: Case $h_1=0$, $h_0=Q_0=1$. Right: $h_0=h_1=Q_0=1$. }}
  \label{fig:OneScalar}
\end{figure}

Let us now consider the case of a one scalar field model. Such a model was introduced in Section \ref{sec:onescalarfield}. As above, we will work with a different but equivalent one scalar model given in Appendix \ref{sec:A}, see 
\eqref{actuppsi}.  We will 
search for non-canonical numerical Lax pairs for (a) the restricted case $h_1=0$ and (b) the general case  corresponding to $h_0, h_1, Q_0 \neq 0$, respectively. An analytic Lax pair for case (a) is given in \eqref{eq:138} 
in Appendix \ref{sec:A}. Accordingly, we set the dimensionality of the numerical Lax pairs to be $6\times 6$. Recall that in order to demonstrate the integrability of the one scalar model, we need to find three integrals of motion. These have been analytically obtained for the $h_1=0$ case as given in \eqref{eq:currentsh0}. 
The ML results are displayed in Figure \ref{fig:OneScalar}. We observe that in both cases the loss decreases down to $10^{-5}$ after training. The comparable losses in both cases suggest that the model with $h_1 \neq 0$ is also integrable, just as the $h_1=0$ case.

The ML experiments given above illustrate that it is possible to train NNs to minimise the loss function corresponding to the Lax pair condition. However, given that the output of these experiments consists of numerical Lax pairs with non zero losses, one must develop as sharp criteria as possible with regards to the necessary condition for the integrability of the system. Developing these integrability criteria would be all the more feasible, if the ambiguity in specifying the target Lax pairs is reduced.  
 Hence, 
 we will now use a different ML approach,
 wherein we stipulate that the Lax pair be in the canonical gauge as defined in \eqref{laxaction}.  
Proving the existence of such a Lax pair fulfills both the necessary and sufficient conditions to establish integrability.

\begin{figure}[h!]
\centering
\includegraphics[scale=.37]{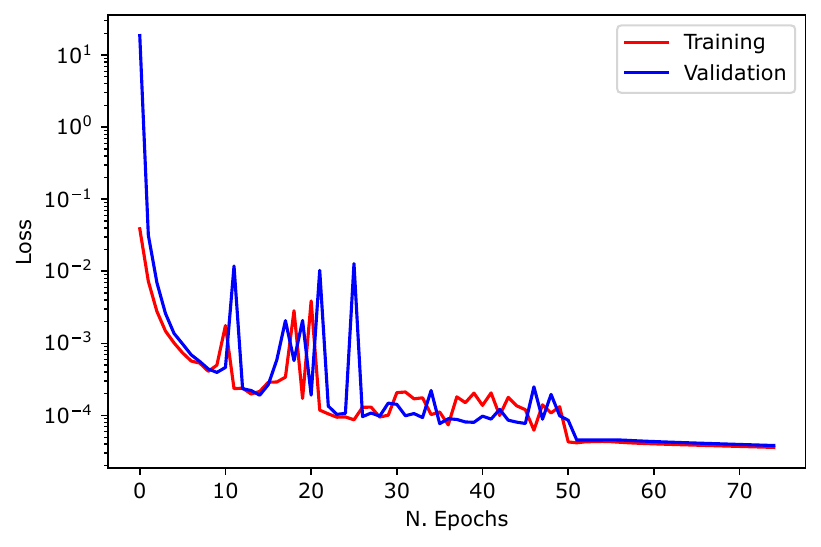}
 \includegraphics[scale=.37]{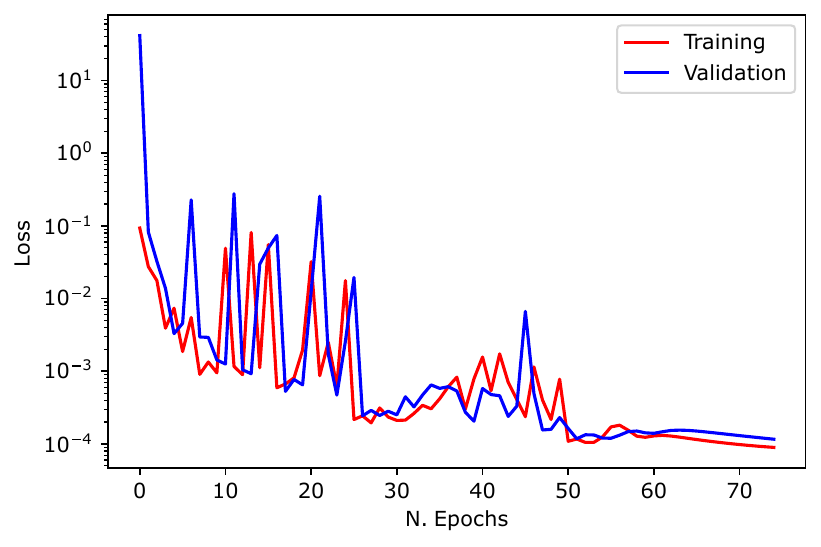}
  \includegraphics[scale=.37]{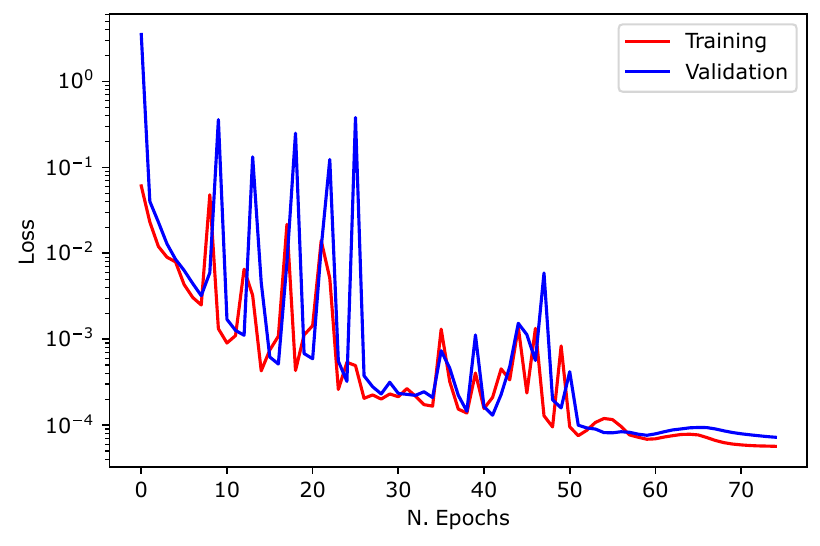}
\caption{\small\textit{ML training of diagonal matrix $L$ in the no scalar field model described by 
\eqref{eq:rescaled}.
Left: Case (a) $h_0=0$, $Q_0=1$. Center: Case (b) $h_0=1$, $Q_0=0$. Right: Case (c) $h_0=Q_0=1$.
}}
  \label{fig:1DModelsD}
\end{figure}

\begin{figure}[h!]
\centering
\includegraphics[scale=.37]{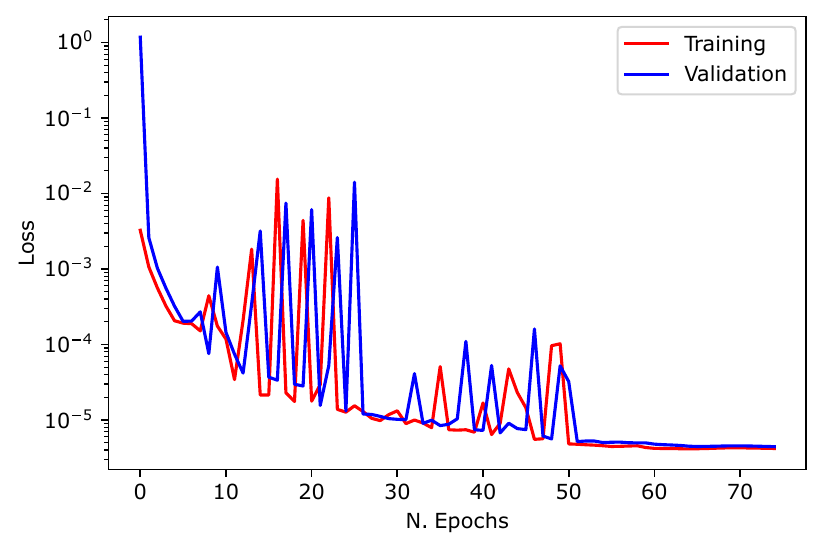}
 \includegraphics[scale=.37]{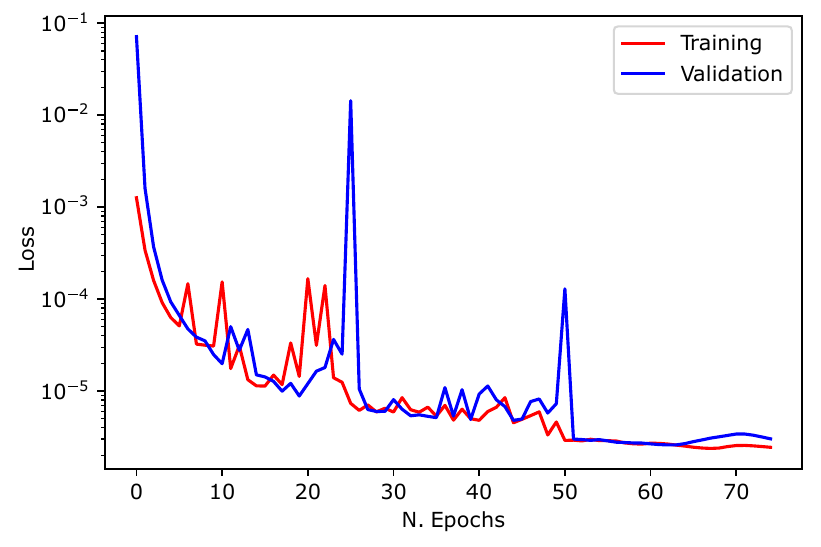}
\caption{\small\textit{ML training of diagonal matrix $L$ in the one scalar field model (see 
\eqref{actuppsi}).
Left: Case $h_1=0$, $h_0=Q_0=1$. Right: $h_0=h_1=Q_0=1$. }}
  \label{fig:OneScalarD}
\end{figure}

We now perform ML searches for diagonal Lax matrices $L$. In the no scalar model we ML search for a diagonal $2 \times 2$ matrix $L$, while in the one scalar model
we ML search for a diagonal $3 \times 3$ matrix $L$. We proceed as in the ML search for numerical Lax matrices described above, using the following loss function, 
\bea
\mathcal{L}=\sum_{i}|(dL_{ii}/dt)|^2+\frac{\alpha_R}{\sum_{i,k}|dL_{ii}/dX_k|^2} \;,
\eea
where we added the regulator term to prevent generating trivial solutions for $L$.
 We have set $\alpha_R=0.1$ in our ML experiments. 
The ML results for the two models are presented in Figures \ref{fig:1DModelsD} and \ref{fig:OneScalarD}, respectively. Here, as in the ML searches discussed above, we observe that in all cases the losses after training reach comparable values.

Having obtained evidence of Liouville integrability we will now proceed to implement a symbolically interpretable ML approach aimed at determining on-shell conserved quantities 
for the various systems under consideration. 
Our approach is to construct a generic conserved quantity as an NN output, symbolically expressed in a chosen basis of phase space variables, then work out its independent components and ultimately verify that these do Poisson commute with each other.

  \section{Machine learning of conserved currents}\label{sec:in}

We now proceed to implement our ML approach to determining currents for some of the 1D systems studied in Section \ref{sec:scla1D}.

In order to train a neural network to search for conserved currents for a given 1D system, we start by defining a set $\mathcal{B}=\{b_i\}$ constituted of functions of the phase space variables. Then for a given degree $k\, \in\, \mathbb{N}$, we define a monomial basis constructed from the elements of $\mathcal B$, 
\bea
\mathcal{B}^{(k)}=\{\prod_{i} b_{i}^{n_i}\}_{0\leq n_i \leq k}\,.
\eea 
For example, in a system with one degree of freedom one could consider $\mathcal{B}=\{p,q\}$ and the monomial basis at degree two would be 
$\mathcal{B}^{(2)}=\{1,p,q,p^2,pq,q^2,pq^2,qp^2,p^2q^2\}$.  
Denoting the monomials in $\mathcal{B}^{(k)}$ as $p_m=\prod_{i} b_{i}^{n_i}$, we can use this set to write a putative current in terms of a single layer linear network,
\begin{equation}\label{eq:LNN}
J_{NN}^{(k)}=w_m p_m\,, \quad p_m\in \mathcal{B}^{(k)},    
\end{equation}
 and train to optimise the weights $w_m$ to minimise the loss function
 \bea \label{Current1}
\mathcal{L}=|dJ_{NN}/dt|^2+\alpha_R\left(\sum_m|dJ_{NN}/d p_m|^2-1\right)^2,
\eea
The term proportional to $\alpha_R$ is a regulator that eliminates phase space independent constants. At each degree $k$, this procedure is essentially a polynomial regression, a best fit algorithm in the space of polynomial functions of degree $k$. Once we have obtained all the predicted on-shell conserved quantities up to a given $k$, we need to extract from them the largest set of currents with mutually vanishing Poisson brackets, as this will form the set $J$ of conserved currents\footnote{For notational simplicity, we use $J$ interchangeably for both the derived ML output as well as for its individual elements.} which, along with the Hamiltonian, are candidates for the diagonal entries of $L$ in the canonical frame. Note that this is ultimately a numerical approximation scheme and hence will generically only serve to indicate the functional form of the putative conserved current. So, at a given $k$, the optimisation might yield a $J_{NN}$ with a few weights being orders of magnitude larger than the rest, indicating that the true $J$ element in this case might simply be expressed in terms of the monomials corresponding to the large weights with the remaining terms set to zero. This drastically constrains the functional search space in which to look for conserved currents, and can potentially serve as a valuable guide for analytic constructions of Lax pairs.  

\begin{figure}[h!]
\centering
\includegraphics[scale=.6]{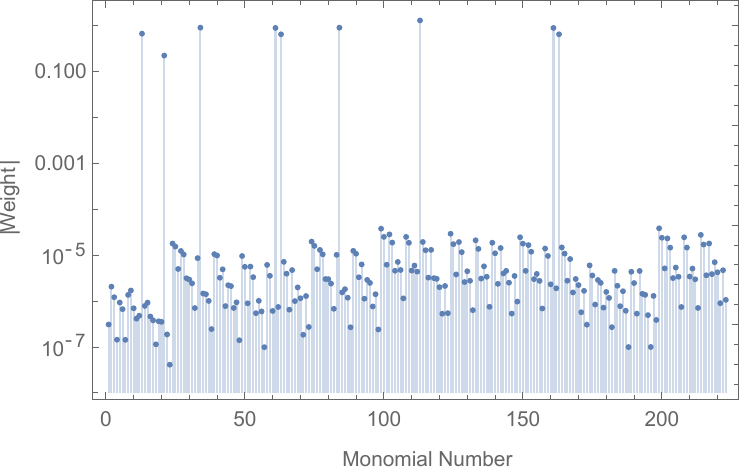}\,\,\,\,\,\,\,\,
 \includegraphics[scale=.6]{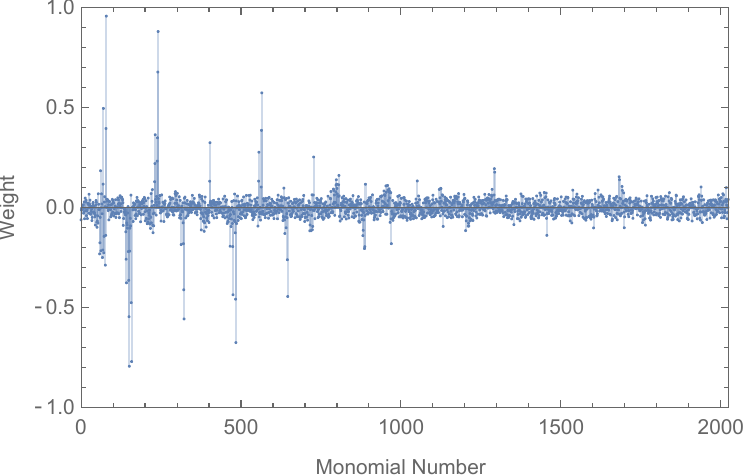}
\caption{\small\textit{No scalar model. Left: Absolute value of the weights for degree two monomials, one sees a clear gap separating the current components from the noise. Right: The value of the trained weights at degree four. In both graphs the basis of monomials was chosen in terms of the fields $U$ and $\psi$, given in \eqref{ac1Dqh}.}}
  \label{fig:LinearNetNoSc}
\end{figure}

 For the no scalar model in Section \ref{sec:scla1D}, given by \eqref{ac1Dqh} with two degrees of freedom $U$ and $\psi$, $J$ is a real function. 
  Here we choose a different basis where the Hamiltonian too can be written as a polynomial in that basis. By looking at the action \eqref{ac1Dqh} we note that the set 
 \bea 
 \mathcal{B}=\{U^\prime,\psi^\prime,e^U,e^\psi,e^{-U},e^{-\psi}\}
 \label{Bps}
 \eea
 can be used to reconstruct all monomials involved in the Hamiltonian at degree $k=2$. There are 224 such monomials after removing the constant term, which would not yield an independent integral of motion. The linear network at degree $k=2$ is then expressed as
 \begin{eqnarray}
     J_{NN}^{(2)}&=& w_1\, U^\prime +w_2\, \psi^\prime+ w_3\, e^U+w_4 \,e^\psi+w_5\, e^{-U}+w_6 \, e^{-\psi}+\ldots +w_{224}(U^\prime \psi^\prime)^2\, .   
 \end{eqnarray}
The ML results for this linear NN applied to the no scalar model are presented in Figure \ref{fig:LinearNetNoSc}, where we have chosen $\alpha_R=1$ in the loss function \eqref{Current1}. There we observe spikes that single out 9 monomial terms from a total of 224. These are the elements with the highest weighted contribution to the ML current, and constitute a constrained set of monomials within which we can search for conserved currents. This search suggests two currents that can be constructed at degree two, which we recognise as the current $P_A$ obtained analytically in Section \ref{sec:nosc}, as well as the Hamiltonian. Further, one can explore the currents appearing at degree $k=4$, where we obtain 2024 monomials. Performing the same experiment as before produces a set of monomials that allow for the construction of $\mathcal H$ and $P_A$ again, as well as products and powers thereof. 

The ML approach so far has served to underline pre-existing knowledge of integrability in the systems under consideration. However this approach is powerful enough to go beyond simply verifying integrability to potentially uncovering the presence of integrable structures in 1D systems. 
We illustrate this by considering the one scalar model \eqref{eq:eqonef}
in the generic case with all three parameters $(Q_0, h_0, h_1)$ turned on.

\begin{figure}
\centering
 \includegraphics[scale=.7]{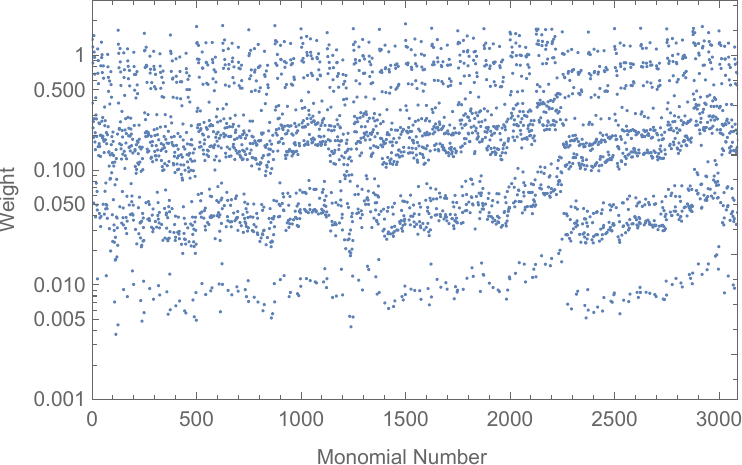}
\caption{\small\textit{Absolute value of the weights for degree 2 monomials in the one scalar model with $h_0,h_1,Q_0\neq 0$.}}
  \label{fig:LinearNetSc3}
\end{figure}

For the one scalar field model we follow a similar approach to the zero scalar case that consists in choosing a basis analogous to \eqref{Bps}, that also includes exponential terms as well as derivatives of the scalar field $\phi$,
\bea
\mathcal{B} =\{U^\prime, \phi^\prime, \psi^\prime, e^U,e^\phi,e^\psi,e^{-U},e^{-\phi},e^{-\psi}\} 
\label{Bexp}
\eea
At degree $k=2$ this basis produces 3374 inequivalent monomials. The optimised weights obtained after training are plotted in Figure \ref{fig:LinearNetSc3}. In contrast to previous experiments we notice a stratified structure, as opposed to a clearly identifiable gap structure that could suggest a conserved current from a smaller set of monomials. Nevertheless, by restricting ourselves to monomials with weights above a chosen cutoff, enables us to reconstruct the Hamiltonian, but not the current \eqref{picurabc}.
This is due to the fact that that the expression for this current involves an integral which can not be accounted for by the linear network \eqref{eq:LNN}.

\begin{figure}
\centering
\includegraphics[scale=.6]{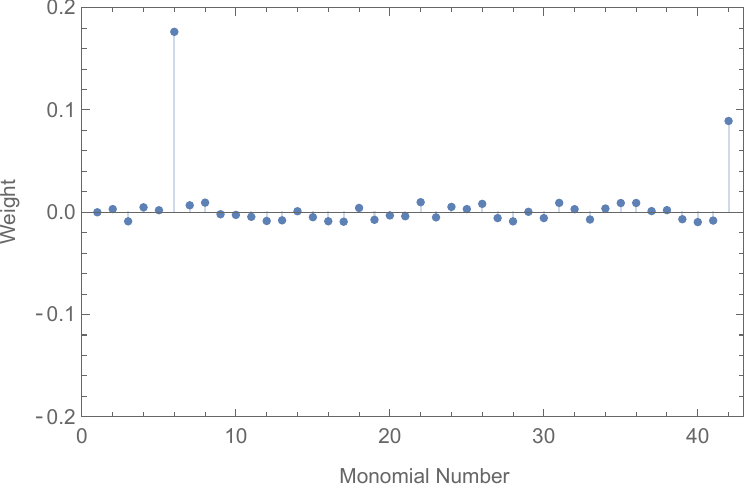}\,\,\,\,\,\,\,\,
 \includegraphics[scale=.6]{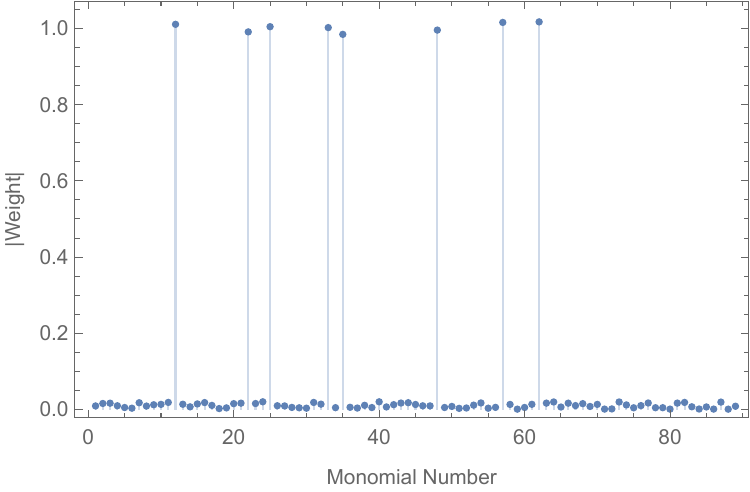}
\caption{\small\textit{Left: Learning of the current $P_C$ for the case when $h_0=0$ in the one scalar model \eqref{abcQh0h1}. Right: Learning of the current $\Pi$ given in \eqref{picurabc}
in the one scalar field model with with $h_0=h_1=Q_0=1$. Both experiments are performed with degree two monomials.}}
  \label{fig:LinearNetNoSc2}
\end{figure}

When $h_0 =0$ and $h_1, Q_1 \neq 0$, to compare with results in Section \ref{sec:onescalarfield}, we use \eqref{basisc} and switch from \eqref{Bexp} to the following set
\bea\label{eq;bs}
\mathcal{B}=\{A,B,C,P_A,P_B,P_C\} \;,
\eea
which at degree $k=2$ leads to 42 monomials. The ML search for  $J_{NN}$ is now implemented using the loss function  \eqref{Current1} with $\alpha_R=1$.
The ML result is displayed on the left hand side of Figure \ref{fig:LinearNetNoSc2}.  We observe two spikes in the plot, which we identify as $P_C$ and $P_C^2$. Inspection of  \eqref{Pabcpr} shows that $P_C$ is indeed a conserved current on-shell.

We now change strategy and perform an ML search for $dJ_{NN}/dt=0$ rather than for $J_{NN}$, aimed at picking up additional currents including those of the form \eqref{picurabc}.
We restrict our discussion to a monomial basis at degree $k=2$. For the one scalar model with $h_0,\,h_1,\,Q_0\neq 0$, we enlarge the set \eqref{eq;bs} by
\begin{equation}\label{eqbasis}
    \mathcal{B}=\{A, B, e^C, A^\prime, B^\prime, 1/B, P_A, P_B, P_C,P_A^\prime, P_B^\prime, P_C^\prime\} \;,
\end{equation}
to obtain 89 monomials of at most degree two in two variables.
The ML search for  $dJ_{NN}/dt=0$ is implemented using the loss function  \eqref{Current1} with $\alpha_R=0$. The ML result is displayed on the right hand side of Figure \ref{fig:LinearNetNoSc2}.  We observe eight spikes of comparable weight ($\approx 1$) corresponding to the monomials $P_C^\prime, A P_A^\prime, B P_B^\prime, A' P_A, B' P_B, 1/B^2, e^{2C}$, $B^2$,
from which one can infer a linear combination which, upon integration, yields the conserved current \eqref{picurabc}.

Finally, we consider a setting in which we ML numerical approximations to conserved quantities $J_{NN}$. The setting is similar to the one described in Section \ref{sec:nums}. We consider a deep neural network with the same architecture as before, but with a single scalar output. 
As a loss function we take 
 \bea 
\mathcal{L}=|dJ_{NN}/dt|^2+ \left(\sum_j|dJ_{NN}/d X_j|^2-1\right)^2,
\eea
where the sum over $j$ is over all the components of the input vector $X$ given in \eqref{XXp}. 
If we consider the no scalar model, following a training process similar to the one described for the Lax pair, we obtain losses of the order $10^{-6}$ after training for the numerical neural network current. Further, knowing that the currents for this system are $\mathcal H$ and $P_A$, we perform a polynomial regression.
This procedure yields the expression, 
\bea
J_{NN}\sim 0.14867 + 0.25717 {\mathcal H}  + 0.24235 P_A
\eea
with a fit accuracy of 92\% over the whole data set.

We have also performed a similar experiment for the one scalar field model with $h_0 = 0$, applying a similar regression procedure. Here we know that $\mathcal{H}$ and $P_C$ are conserved currents for the system, obtained as  
\bea
J_{NN}\sim -2.11911 - 0.77678 \mathcal{H} + 0.00381 \mathcal{H} P_C
\eea
with a fit accuracy of 40\% only. This low accuracy indicates that the basis of currents $\mathcal{H}$, $P_C$ for our polynomial regression is incomplete.

\section{Conclusions \label{sec:conc}}

In this paper we have identified integrability structures in 2D gravitational theories resulting from a 
2-step reduction of certain 4D gravitational theories describing the coupling of Maxwell fields and neutral scalar fields $\phi$ to gravity in the presence of a scalar potential $V(\phi)$. Restricting ourselves to the subspace of solutions given in \eqref{eq:rhot}, we have demonstrated that a subset of the 2D equations of motion can be viewed as compatibility equations for a linear system, which in turn can be represented in terms of differential operators $({\cal L}, {\cal M})$, and which reduces to the BM system in the vanishing scalar potential case.  Our 2D  integrability analysis has 2 caveats. Firstly, it can only be applied to solutions which continue to exist as the scalar potential is taken to zero. Secondly, the linear system encodes only part of the equations of motion and guarantees only partial integrability. 

Recognising that our chosen solution subspace is amenable to a 1D reduction, both these caveats can be circumvented by a complementary analysis of this subspace from a 1D point of view.  For this purpose we consider an effective 1D theory whose solution space contains the class of solutions defined by $\mathcal{H}=0$ which correspond to the chosen subspace of 4D gravitational configurations.

Subsequently we analyse  Liouville integrability realised by constructing a Lax pair of matrices  
$(L,M)$  in these models. We adopt a canonical frame wherein we can choose $M$ to be an identity matrix of dimension equal to that of the configuration space, while $L$ is a diagonal matrix.  The Lax equation then renders the diagonal elements to be constants of motion.  We choose the Hamiltonian to be one of the diagonal elements. Only for an integrable system can we write down the $L$ matrix with  the diagonal filled up by independent conserved currents including the Hamiltonian. That is, a system is integrable iff such a $L$ matrix exists. 
This effectively implies that the 1D criterion for integrability of the 4D solution subspace under consideration reduces to determining the number of independent Poisson commuting conserved currents.

Our results are constituted of  ML driven numerical searches for Lax pairs as well as their analytic constructions, both canonical and generic, as well as construction of conserved currents as functions of phase space variables, by both analytic techniques as well as symbolic regression in ML. 
We find that the ML implementation of numerical Lax searches in a given model can act as a indicator of integrability, when comparing the loss function value for this search with that of a Lax search in a system known to be integrable. The symbolic regression technique offers a framework to systematically detect integrability through verifying the existence or lack thereof of conserved currents in the chosen basis of phase space variables.

To sum up, in this note, we have illustrated, with examples, a proof-of-concept point about how ML approaches could be implemented effectively in the study of integrable structures. Our ML algorithm incorporates broad non-model dependent physics principles that help transcode the problem of searching for matrix Lax pairs to functional regression problems, rendering the approach applicable to systems whose integrability structures are not known a priori.
Hence, as opposed to prior literature \cite{Ishikawa_2021,Krippendorf:2021lee,daigavane2022learning,Lal:2023dkj}
in this subject, which has used ML techniques to validate analytically known integrable structures, our ML approach does not use an a priori knowledge of the integrability of the system and augments our analytic search for integrals of motion.

\section*{Acknowledgements}
We would like to thank Giorgi Butbaia, Gianluca Inverso, Yang-Hui He, Vishnu Jejjala, Axel Kleinschmidt, Shailesh Lal, Swapna Mahapatra, Suvajit Majumdar, Challenger Mishra, Thomas Mohaupt, Silvia Nagy, Justin Tan and Alessandro Torrielli for helpful discussions.
The authors are supported
by FCT/Portugal through 
through project UIDB/04459/2020 with DOI identifier 10-54499/UIDP/04459/2020.
The authors would like to thank the Isaac Newton Institute for Mathematical Sciences for support and hospitality during the program {\it Black holes: bridges between number theory and holographic quantum information} when work on this paper was undertaken; this work was supported by EPSRC grant number EP/R014604/1.

\appendix

\section{The field equations in two dimensions  \label{sec:eomsL2}}

We consider the two-dimensional Lagrangian \eqref{Lag2d} and derive the associated equations of motion using standard techniques. 
For the benefit of the reader, we work out the various steps involved.

 First we consider the equations of motion for the scalar fields encoded in $M$.
We will denote these scalar fields by $\Phi^I$; they include the scalar fields $\Delta, \phi$.
Using 
 \bea
 \delta A_{\mu} &=& - M^{-1} \delta M \, A_{\mu} + M^{-1} \partial_{\mu} \delta M \nonumber\\
 &=& - M^{-1} \delta M \, A_{\mu} + \partial_{\mu} \left( M^{-1} \delta M \right) + A_{\mu} \, M^{-1} \delta M, 
 \eea
 we obtain
 \bea
 \frac12 \Tr \left[ M^{-1} \frac{\delta M}{\delta \Phi^I}  \, \nabla^{\mu} \left( {\tilde \rho} A_{\mu} \right) \right] - \frac{\delta {\cal V}}{\delta \Phi^I}  = 0 .
 \label{eomPhi}
 \eea
 Now we use the fact that $M^{-1} \frac{\delta M}{\delta \Phi^I}$ is an element of the Lie algebra $\mathfrak{p} $, whose generators we denote by $D^I$.
 Note that the $D^I$ are constant matrices, and that their number equals the number of fields $\Phi^I$.
 We therefore have
 \bea
 M^{-1} \frac{\delta M}{\delta \Phi^I} &=& f_{IJ} (\Phi) \, D^J \;, \nonumber\\
 A_{\mu} &=& \partial_{\mu} \Phi^I  \, f_{IJ} (\Phi) \, D^J \;.
 \label{relAD}
 \eea
 Then, the equation of motion \eqref{eomPhi} can be written as
 \bea
 \frac12 \Tr \left[ D^K D^L \right]    f_{I K} (\Phi) \, \nabla^{\mu} \left(  {\tilde \rho} \, \partial_{\mu} \Phi^J  \, f_{J L} (\Phi) \right) 
 - \frac{\delta {\cal V}}{\delta \Phi^I}  = 0 .
 \eea
 Denoting
 \bea
 C^{K L} &=& \frac12 \Tr \left[ D^K D^L \right] = C^{LK}\;, \nonumber\\
 T_I{}^L &=&  f_{I K} (\Phi) \, C^{KL} \;,
 \eea
 and assuming that $ T_I{}^L $ is invertible with inverse $(T^{-1})_J{}^I$ , i.e. $ (T^{-1})_J{}^I T_I{}^L = \delta_J^L$, we obtain
 \bea
 \nabla^{\mu} \left(  {\tilde \rho} \, \partial_{\mu} \Phi^J  \, f_{J L} (\Phi) \right) -  (T^{-1})_L{}^I \, \frac{\delta  {\cal V}}{\delta \Phi^I }= 0 \;.
 \eea
 Contracting this equation with $D^L$ gives 
 \bea
 \nabla^{\mu} \left( {\tilde \rho}  A_{\mu} \right) = {\tilde \rho} \, G, 
 \label{eomM}
  \eea
 where the function $G$ is given by
 \bea
\label{rR}
{\tilde \rho} \, G=   D^L \,  (T^{-1})_L{}^I \, \frac{\delta {\cal V} }{\delta \Phi^I } \;.
 \eea
 Using \eqref{relAD}, we establish the relation
 \bea
  \frac12  \,   \Tr \left[ D^L \,  (T^{-1})_L{}^I \, \frac{\delta {\cal V} }{\delta \Phi^I } \, A_{\mu} \right] = \frac{\delta {\cal V} }{\delta \Phi^I } \partial_{\mu} \Phi^I \;.
  \label{relVA}
  \eea
Taking $ds_2^2$ to be a flat metric given in \eqref{flat2dm},
and using \eqref{stho}, the field equation \eqref{eomM} can be written as
  \bea
 \label{eomAR}
 d \left( \tilde{\rho} \star A \right) = dv \wedge d\rho \, {\tilde \rho} \left( \lambda \, \partial_{\rho} A_{\rho} + \partial_v A_v + \lambda \, 
\frac{\partial {\tilde \rho}}{\partial \rho} \, \frac{ A_{\rho}}{\tilde \rho}  + 
\frac{\partial {\tilde \rho}}{\partial v} \, \frac{ A_{v}}{\tilde \rho} 
 \right) = 
dv \wedge d\rho  \, {\tilde \rho} \, G \;\;\;,\;\;\; A = M^{-1} d M .\nonumber\\
\eea

Next, varying the Lagrangian \eqref{Lag2d} with respect to 
$2\Sigma,  {\tilde \rho}, g_{\mu \nu}$ gives the equations of motion
\bea
&& \Box {\tilde \rho} = - \frac{\delta {\cal V}}{\delta (2\Sigma)} = - {\cal V} , 
 \nonumber\\
 && \Box (2 \Sigma ) = R_2 -  \frac14 g^{\mu \nu} \Tr \left( A_{\mu} A_{\nu} \right)   - \frac{\delta {\cal V}}{\delta {\tilde \rho}} = 
R_2 -  \frac14 g^{\mu \nu} \Tr \left( A_{\mu} A_{\nu} \right)   - \frac{{\cal V} }{\tilde \rho}, 
\nonumber\\
 &&\partial_{\mu} {\tilde \rho} \, \partial_{\nu} (2 \Sigma) +  \partial_{\nu} {\tilde \rho} \, \partial_{\mu} ( 2 \Sigma) - 
 g_{\mu \nu} \, \partial_{\sigma} {\tilde \rho} \,  \partial^{\sigma} ( 2 \Sigma) 
 + 2 g_{\mu \nu} \, \Box {\tilde \rho} - 2 \nabla_{\mu} \nabla_{\nu} {\tilde \rho} \nonumber\\
 && - \frac12 {\tilde \rho} \left( \Tr A_{\mu} A_{\nu} - \frac12 g_{\mu \nu} \Tr A_{\sigma} A^{\sigma} \right)
   + g_{\mu \nu} {\cal V} = 0 .
 \label{eomA}
\eea

Taking  $ds_2^2$ to be the flat metric given in \eqref{flat2dm}, these field equations become
 \bea
 \label{fieldeom}
&& \lambda \partial_{\rho}^2 {\tilde \rho} + \partial_{v}^2 {\tilde \rho} = -  {\cal V} , 
 \nonumber\\
 &&  \lambda \partial_{\rho}^2 (2 \Sigma )  + \partial_{v}^2  (2 \Sigma ) = 
 -  \frac14 \Tr \left( \lambda \, A_{\rho} A_{\rho } + A_v A_v \right)   -  \frac{{\cal V} }{\tilde \rho} , \nonumber\\
   &&(\mu, \nu) = (\rho, \rho): \quad \partial_{\rho} {\tilde \rho} \, \partial_{\rho} (2 \Sigma) - \lambda  \partial_{v} {\tilde \rho} \, \partial_{v} (2 \Sigma) + 2 \lambda 
  \partial_v^2  {\tilde \rho} 
- \frac14 \, {\tilde \rho}  \,   \Tr \left[ A_{\rho} A_{\rho} - \lambda A_v A_v \right] + \lambda \, {\cal V} = 0 , \nonumber\\
&&(\mu, \nu) = (v, v): \quad 
\partial_{v} {\tilde \rho} \, \partial_{v} (2 \Sigma) - \lambda  \partial_{\rho} {\tilde \rho} \, \partial_{\rho} (2 \Sigma) + 2 \lambda 
  \partial_{\rho}^2  {\tilde \rho} 
- \frac14 \, {\tilde \rho}  \,   \Tr \left[ A_{v} A_{v} - \lambda A_{\rho}  A_{\rho} \right] +  \, {\cal V} = 0
\;, \nonumber\\
&&(\mu, \nu) = (\rho, v): \quad \partial_{\rho} {\tilde \rho} \, \partial_{v} (2 \Sigma) + \partial_{v} {\tilde \rho} \, \partial_{\rho} (2 \Sigma) - 2 \partial_{\rho} \partial_v 
{\tilde \rho} - \frac12  \, {\tilde \rho}  \,   \Tr \left[ A_{\rho} A_{v} \right] = 0 \;.
\eea
Combining the third equation with the fourth equation gives the first equation. We may therefore take the first, second, third and fifth equations as independent
equations.

Next, we analyse the solvability of the field equations \eqref{fieldeom}.
Taking the $\rho$-derivative of the third equation and adding it to $\lambda$ times the $v$-derivative of the fifth equation yields
\bea
 \partial_{\rho} \left( \log {\tilde \rho} + 2 \Sigma \right) {\cal V} &=& \partial_{\rho} {\cal V} - \frac12 {\tilde \rho}  \,   \Tr \left[ G \, A_{\rho} \right] =
  \partial_{\rho} {\cal V} - \frac12  \,   \Tr \left[ D^L \,  (T^{-1})_L{}^I \, \frac{\delta {\cal V} }{\delta \Phi^I } \, A_{\rho} \right] \nonumber\\
  &=& 
   \partial_{\rho} {\cal V} - 
    \frac{\delta {\cal V} }{\delta \Phi^I } \partial_{\rho} \Phi^I  \;,
    \label{solvr}
 \eea
 where we used the first two field equations given in \eqref{fieldeom} as well as \eqref{eomAR} and 
 \eqref{relVA}. On the other hand, taking the $v$-derivative of the third equation and subtracting the $\rho$-derivative of the fifth equation gives
\bea
 \partial_{v} \left( \log {\tilde \rho} + 2 \Sigma \right) {\cal V} &=& \partial_{v} {\cal V} - \frac12 {\tilde \rho}  \,   \Tr \left[ G \, A_v \right] =
  \partial_{v} {\cal V} - \frac12  \,   \Tr \left[ D^L \,  (T^{-1})_L{}^I \, \frac{\delta {\cal V} }{\delta \Phi^I } \, A_{v} \right] \nonumber\\
  &=& 
   \partial_{v} {\cal V} - 
    \frac{\delta {\cal V} }{\delta \Phi^I } \partial_{v} \Phi^I  \;,
     \label{solvv}
 \eea
 where we used the first two field equations given in \eqref{fieldeom} as well as \eqref{eomAR} and 
 \eqref{relVA}.
 
 We note that equations \eqref{solvr} and \eqref{solvv} 
are automatically satisfied, since they can be written as 
 \bea
\partial_{\mu} \left( \log {\tilde \rho} + 2 \Sigma \right) {\cal V} &=&   \partial_{\mu} {\cal V} - 
    \frac{\delta {\cal V} }{\delta \Phi^I } \partial_{\mu} \Phi^I  \;.
    \label{conscond2}
\eea
Here we have made use of the fact that since $\cal V$ has the form given in \eqref{calvpot}, we have
\bea
\partial_{\mu} {\cal V} - 
    \frac{\delta {\cal V} }{\delta \Phi^I } \partial_{\mu} \Phi^I  
    = \partial_{\mu}\left( \log {\tilde \rho} + 2 \Sigma \right) {\cal V} \;.
\eea

Summarizing, the independent field equations are
\bea
d \left( \tilde{\rho} \star A \right) = dv \wedge d\rho  \, {\tilde \rho} \, G \;\;\;,\;\;\; A = M^{-1} \, d M 
\eea
together with
\bea
&& \lambda \partial_{\rho}^2 {\tilde \rho} + \partial_{v}^2 {\tilde \rho} = - {\cal V} , \nonumber\\
&&  \lambda \partial_{\rho}^2 (2 \Sigma )  + \partial_{v}^2  (2 \Sigma ) = 
 -  \frac14 \Tr \left( \lambda \, A_{\rho} A_{\rho } + A_v A_v \right)   -  \frac{{\cal V} }{\tilde \rho} , \nonumber\\ 
  &&(\mu, \nu) = (\rho, \rho): \quad \partial_{\rho} {\tilde \rho} \, \partial_{\rho} (2 \Sigma) - \lambda  \partial_{v} {\tilde \rho} \, \partial_{v} (2 \Sigma) + 2 \lambda 
  \partial_v^2  {\tilde \rho} 
- \frac14 \, {\tilde \rho}  \,   \Tr \left[ A_{\rho} A_{\rho} - \lambda A_v A_v \right] + \lambda \, {\cal V} = 0 , \nonumber\\
&&(\mu, \nu) = (\rho, v): \quad \partial_{\rho} {\tilde \rho} \, \partial_{v} (2 \Sigma) + \partial_{v} {\tilde \rho} \, \partial_{\rho} (2 \Sigma) - 2 \partial_{\rho} \partial_v 
{\tilde \rho} - \frac12  \, {\tilde \rho}  \,   \Tr \left[ A_{\rho} A_{v} \right] = 0 \;.
\eea

\section{Families of Lax pairs for 1D models \label{sec:A}}

A 1D system described by the Hamiltonian  
\begin{equation}\label{eq:hamiltonian1D}
    \mathcal{H} (p,q)=\frac12 p^2+V(q)
\end{equation}
admits several 
Lax pairs $(L,M)$, i.e. pairs $(L,M)$ satisfying
\bea
\frac{d}{dt} L = [L, M] 
\label{eq:LaxPairEq}
\eea
on-shell.
Consider the following non-canonical pair
\begin{equation}
L = 
\begin{pmatrix}
p  & \frac{h(p,q) f(q)}{\lambda}\\
\frac{\lambda f(q)}{h(p,q)} & - p 
\end{pmatrix} ,\;\;\; M = 
\begin{pmatrix}
a & \, \frac{c \, h^2(p,q)}{\lambda^2}+ \frac{h(p,q) f^\prime (q)}{2\lambda} \\
c-\frac{\lambda f^\prime (q)}{2 h(p,q)} & a-\frac{2 c \,  p \,  h(p,q)}{ \lambda \, f(q)}+\frac{p \, 
\partial_q h(p,q)}{h(p,q)}-\frac{f(q)f^\prime(q) \partial_p h(p,q)}{h(p,q)} 
\end{pmatrix}\,,   
\label{LMac}
\end{equation}
where $a$ and $c$ are functions of $p$ and $q$ and where $\lambda\in\mathbb{R}\backslash\{0\}$ is a parameter. One can show that this pair satisfies \eqref{eq:LaxPairEq} by virtue of Hamilton's equations $\dot{q} = \partial \mathcal{H}/\partial p= p$ and $\dot{p} = - \partial {\cal H}/\partial q = - \partial V / \partial q$. Thus, regardless of the particular choice of the functions $a$ and $c$, this pair $(L,M)$ constitutes a Lax pair that also 
depends on the parameter $\lambda$. 
Furthermore we note that
\begin{equation}
\label{trL2H}
    {\rm Tr}(L^2)=4 \mathcal{H} \;,
\end{equation}
provided that $f$ is expressed in terms of the potential $V$ by 
\bea
f^2 (q) = 2 V (q) \,,
\eea
which we will assume to be the case in the following.

Next, let us pick specific choices for the functions $a$ and $c$.
First, let us choose $a=c=0$ and let us set $h(p,q)=1$, in which case \eqref{LMac}
becomes
\begin{align}
{\rm FAMILY}\,\,{\rm I:}& \quad L = 
\begin{pmatrix}
p  & \frac{f(q)}{\lambda}\\
\lambda f(q) & - p 
\end{pmatrix} ,\;\;\; M = 
\begin{pmatrix}
 0 & \, \frac{ f^\prime (q)}{2 \lambda} \\
-\frac{\lambda f^\prime (q)}{2} & 0 
\end{pmatrix}\,,\label{eq:f1}
\end{align}
which describes a family of Lax pairs for the system described by \eqref{eq:hamiltonian1D}. This family is only defined for positive $V$.  A different family of Lax pairs, which is valid also for $V<0$, is obtained by choosing
$a=0$, $c=\frac{\lambda f^\prime (q)}{2 h(p,q)}$ and $h(p,q)=f(q)$,
\begin{align}
{\rm FAMILY}\,\,{\rm II:}& \quad L = 
\begin{pmatrix}
p  & f^2(q)/\lambda\\
\lambda & - p 
\end{pmatrix} ,\;\;\; M = 
\begin{pmatrix}
 0 & \,\quad (f(q)^2)^\prime/2\lambda \\
0 & 0 
\end{pmatrix}\,.\label{eq:f2}
\end{align}

Next, we will use these two specific families of Lax pairs as a guide to constructing Lax pairs for the systems described by the
Lagrangians discussed in Section \ref{sec:scla1D}.

First, let us consider the action \eqref{ac1Dqh}. By redefining the integration variable as $e^{- 2 \psi (r)} dr = dR $, this action can 
be recast as 
\begin{equation}\label{eq:rescaled}
   -S_{1D}=\int dR \left[(U^\prime)^2-(\psi^\prime)^2+\frac14 Q_0^2 e^{2U}-\frac34 h_0^2 e^{-2U+4\psi}\right]\,.
\end{equation}
We notice that when $h_0=0$, the two fields $U$ and $\psi$ decouple, in which case the action describes two decoupled systems of the form 
$\mathcal{L}= \frac12 \dot{q}^2-V(q)$, each with a description in terms of a Lax pair as described above. Thus, using $f^2 = - \frac14 Q_0^2 e^{2U}$ and choosing $\lambda = -1$, a Lax pair describing these two decoupled systems is
given by
\begin{equation}\label{eq:laxp1}
    L=\begin{pmatrix} U^\prime &  \frac14 Q_0^2 e^{2U} & 0 & 0\\
    -1 & -U^\prime & 0 & 0\\
    0 & 0 & 0 & \psi^\prime \\
    0 & 0 & -\psi^\prime & 0 
    \end{pmatrix}\,, \quad M=\begin{pmatrix} 0 &  \frac14 Q_0^2 e^{2U} & 0 & 0\\
    0 & 0 & 0 & 0\\
    0 & 0 & 0 & 0 \\
    0 & 0 & 0 & 0 
    \end{pmatrix} \;.
\end{equation}
When switching on $h_0$, we expect these $4 \times 4$ matrices to get deformed
by terms proportional to $h_0$. 
However, we were not able to obtain these deformed $4 \times 4$ matrices, and so in Section \ref{sec:nosc}
we instead presented a Lax pair
for the full system based on $3 \times 3$ matrices $L$ and $M$. 

Next, let us consider the action \eqref{eq:eqonef}. 
By redefining the integration variable as $e^{- 2 \psi (r)} dr = dR $, this action can be recast as 
\bea
   -S_{1D}&=&\int dR\left[(U^\prime)^2+(\phi^\prime)^2-(\psi^\prime)^2+\frac12 e^{2U+2\phi} Q_0^2 \right. \nonumber\\
   && \left. \qquad \qquad 
- e^{-2U+4\psi} \left(h_0 h_1 + \frac12 \left(h_0 e^\phi+h_1 e^{-\phi}\right)^2 \right)\right]\,. 
\label{actuppsi}
\eea
Let us now consider three limiting cases of \eqref{actuppsi}, resulting from switching off one of the three parameters $(Q_0, h_0, h_1)$. 

First, let us consider the case when $h_1 = 0$. Redefining the fields as
\begin{eqnarray}
    \mathcal{\mathcal{A}}&=&U+\phi\,\nonumber\\
    \mathcal{B}&=&U-\phi-2\psi\,\nonumber\\
    \mathcal{C}&=&-U+\phi+\psi, 
\end{eqnarray}
the Lagrangian becomes
\bea
\mathcal{L}_{h_1=0}=\frac12\left(\mathcal{A}^{\prime\,2} - \mathcal{B}^{\prime\,2} + 2 \mathcal{C}^{\prime\,2}+ e^{2 \mathcal{A}} Q_0^2- e^{-2 \mathcal{B}} h_0^2 \right) \;.
\eea
The three fields $\mathcal{A},\mathcal{B}$ and $\mathcal{C}$ are thus decoupled, each having a description in terms
of a Lax pair as described above. A Lax pair describing these three decoupled systems
is given by the following $6 \times 6$ matrices,
\bea
\label{eq:138}
{\tiny
L=\begin{pmatrix} \mathcal{A}^\prime &  Q_0^2 e^{2\mathcal{A}} & 0 & 0 & 0 & 0\\
    -1 & -\mathcal{A}^\prime & 0 & 0  & 0 & 0\\
    0 & 0 & h_0 e^{-\mathcal{B}} & \mathcal{B}^\prime  & 0 & 0 \\
    0 & 0 & -\mathcal{B}^\prime & -h_0 e^{-\mathcal{B}}  & 0 & 0 \\
    0 & 0 & 0 & 0  & \sqrt{2} \mathcal{C}^\prime & 0 \\
    0 & 0 & 0 & 0  & 0 & -\sqrt{2} \mathcal{C}^\prime \\
    \end{pmatrix}\,, \quad M=\begin{pmatrix} 0 &  Q_0^2 e^{2\mathcal{A}} & 0 & 0 & 0 & 0\\
    0 & 0 & 0 & 0 & 0 & 0\\
    0 & 0 & 0 & -\frac12 h_0 e^{-\mathcal{B}}& 0 & 0\\
    0 & 0 & -\frac12 h_0 e^{-\mathcal{B}} & 0 & 0 & 0\\
    0 & 0 & 0 & 0 & 0 & 0\\
    0 & 0 & 0 & 0 & 0 & 0
    \end{pmatrix}
  } \,.
  \nonumber\\
  \eea
We note that $L$ satisfies ${\rm Tr}(L^2)=4 \mathcal{H}$. Since the system is decoupled we can write down three independent Poisson commuting  conserved currents 
  \begin{equation}\label{eq:currentsh0}
  J_1=\mathcal{H}\,, \quad J_2=P_\mathcal{C}=2\mathcal{C}^\prime\,,\quad J_3=P_\mathcal{A}^2 -e^{2\mathcal{A}}Q_0^2=(\mathcal{A}^\prime)^2 -e^{2\mathcal{A}}Q_0^2\,.
  \end{equation}
  
Next let us consider the case $h_0=0$. The action \eqref{actuppsi} can be brought to a form similar to \eqref{eq:rescaled} by redefining the fields $U$ and $\phi$ as
$\mathcal{A}=U+\phi$, $\tilde{\mathcal{B}}=U-\phi$. In this basis the action becomes 
\bea
   -S_{1D}=\frac12 \int dR\left[(\mathcal{A}^\prime)^2+(\tilde{\mathcal{B}}^\prime)^2-2(\psi^\prime)^2+ e^{2\mathcal{A}} Q_0^2
- e^{-2\mathcal{A}+4\psi}h_1^2\right]\,. 
\label{acth0}
\eea
In this case we find three 
Poisson commuting  conserved quantities, namely 
\bea
J_1=\mathcal{H} \;\;\;,\;\;\;  J_2 = P_{\tilde{\mathcal{B}}}=\tilde{\mathcal{B}}^\prime\,,\quad J_3=P_{\mathcal{A}}-\int_{+\infty}^{R}  (e^{2\mathcal{A}} Q_0^2
+ e^{-2\mathcal{A}+4\psi}h_1^2) \, d {\tilde R}\;,
\label{J1231}
\eea
where $P_{\mathcal{A}} = \mathcal{A}'$.
Note that the current $J_2$ is precisely the current $P_C$ identified in Section \ref{sec:onescalarfield}.

Finally, let us consider the case $Q_0=0$. Defining $\hat{\mathcal{A}}=2U-\psi$ and $\hat{\mathcal{B}}=U-2\psi$, the action \eqref{actuppsi} becomes 
\bea
   -S_{1D}=\int dR\left[\frac13 ( \hat{\mathcal{A}}^\prime)^2-\frac13 (\hat{\mathcal{B}}^\prime)^2+(\phi^\prime)^2 
- e^{-2\hat{\mathcal{B}}} \left(h_0 h_1 + \frac12 \left(h_0 e^\phi+h_1 e^{-\phi}\right)^2 \right)\right]\,,
\label{actQ}
\eea
from which we infer the following conserved quantities
\begin{equation}
J_1=\mathcal{H}\,\,,\, \, J_2=P_{\hat{\mathcal{A}}}=\frac23\hat{\mathcal{A}}^\prime
\,\, ,\,\, J_3=P_{\hat{\mathcal{B}}}\,-\int_{+\infty}^R  e^{-2\hat{\mathcal{B}}} \left(h_0 h_1 + \frac12 \left(h_0 e^\phi+h_1 e^{-\phi}\right)^2 \right) d {\tilde R} \,,
\label{J1232}
\end{equation}
where $P_{\hat{\mathcal{B}}} = -\frac23 \mathcal{B}'$.

\section{Machine Learning Lax pairs \label{sec:MLp}}

A particular feature of neural networks is their success at making predictions over large data sets. However, reverse engineering the network to obtain what it is actually doing is less trivial. The problem at hand was to build neural networks that can produce Lax pairs for a given system as a first step towards demonstrating their integrability. In order for the network to produce a sensible result, finding an analytic expression is unavoidable. 

If a Lax pair exists one expects the Lax pair equation \eqref{eq:LaxPairEq} to be satisfied identically over the entire phase space. A neural network could then be used to approximate $L$ and $M$ as neural network outputs. However one has to bear in mind the following. First, neural networks are better suited to make predictions over compact domains. Hence we expect a neural network to give a good approximation only over a compact domain that contains the training set. Second, since we do not work at any of the limits where the universal approximation holds, we have to rely on a neural network that is expressive enough (enough layers, enough neurons per layer, suitable activation function), such that upon training, the error in the approximation can be controlled by a parameter $\epsilon$ such that 
\begin{equation}
    \epsilon >\Delta_{ij}= { \displaystyle \max_{(p_\alpha,q_\alpha) }  } \in\mathcal{D} 
    \left|L_{ij}-f^{NN}_{ij}\right|^2 \,,
\end{equation}
 with $f^{NN}_{ij}$ being the NN approximation function for the $ij$ component of $L$, and similarly for $M$. The parameter $\epsilon$  is an architecture dependent quantity and it is supposed to approach zero as one reaches either infinite width or infinite length, provided a Lax pair exists. However, we are limited to work with a finite network, and therefore must look for a 
 suitable NN architecture for our problem, since a systematic architecture analysis is not feasible with the resources at our disposal.

Using the neural network as an approximation for a Lax pair in a system where no analytic Lax pair is known, presents us with the issue of decidability. When no 
analytical expressions for $L$ and $M$ are known, one can not estimate $\epsilon$ for a NN approximation. The only parameter at hand is the loss function which, as in  \eqref{eq:Loss}, can be designed to measure deviations from the Lax equation (with additional regulator terms). After training, the loss is going to be fixed at a certain value different from zero, and therefore we must judge if that value is small enough to ensure the robustness of the numerical approximation. In the following we highlight the basic differences between the numerical Lax pair approximation by NNs and the framing of the actual Lax pair search as an optimisation problem.
\begin{itemize}
    \item If we construct a neural network with traditional activation functions (e.g. ReLU, sigmoid, tanh) the output produced is going to be an analytic expression, given as a composition of activation functions at each network layer and decorated with weights and biases. This analytic expression is then used to generate a \textbf{numerical} Lax pair. A numerical Lax pair is a pair of matrices $L$ and $M$ whose entries are neural network outputs such that after training the Lax loss (i.e. a loss that measures deviations from $L$ and $M$ satisfying the Lax pair equation \eqref{eq:LaxPairEq}) reaches values below a certain threshold value. The result is a pair of matrices with numerical values at each phase space point.  Furthermore, the training is performed over a compact set in phase space, where the loss function takes an acceptably small minimum value. As we move away from this region the NN approximation will become worse. These factors limit the interpretability of the output in terms of deciding whether the system is integrable or not. 
    \item Approximating the Lax pair by neural networks (NNs) has the inherent problem of interpretability: even in the case of small numerical deviations from the Lax pair equation \eqref{eq:LaxPairEq}, the Lax pair approximations are hard to recast into an interpretable form, i.e. into analytic expressions that actually satisfy the Lax pair equation. If we take the existence of the Lax pair as the criterion for integrability of the 1D systems under consideration, one has to convert these numerical values into a meaningful functional output that exactly satisfies the Lax pair equation. 
    
    \item  If after training, an acceptably small loss is reached, this may hint at the integrability of the system: 
    consider a system for which a Lax pair is known; after training one obtains a numerical Lax pair for this system. This output provides a measure of what type of loss is to be expected for the given NN architecture and the type of system under consideration. We now switch on a deformation and train again to obtain a numerical Lax pair for the deformed system. If the loss at the end of the training process is of the same order as the one obtained for the undeformed system, then we consider the deformation to have a good chance of being integrable as well \cite{Krippendorf:2021lee,Lal:2023dkj}. This type of analysis is carried out in Section \ref{sec:nums}. In order to ensure that the deformed system has a chance of  being integrable, the numerical ML output has to be made concrete by casting the numerical Lax pair into an analytic one, valid over the whole phase space. One possible means to achieve this is to apply symbolic regression: some options to try in this direction are packages such as PySR \cite{cranmer2020discovering}. We leave this for future work. 
    \item An alternative way is to use NNs that simulate the evolution of parameters in the physical problem.
    This requires making assumptions, such as what type of functional expressions appear in the entries of the Lax pair, and what is a suitable basis for those terms. We denote these as \textbf{interpretable} Lax pairs. Since we work with suitable non-standard activation functions, this interpretable approach can not be generalised to any other one dimensional Hamiltonian, as the choice of a basis is model dependent. While in this case one trains again over a compact dataset, there is the chance that upon training the resulting approximation functions can be turned into actual analytical results, valid away from the training set. We explore this alternative in Section \ref{sec:in}.
\end{itemize}

At last, let us make a comment about the dimensionality of the Lax pairs. In Appendix \ref{sec:A} we presented two families of $2\times 2$ Lax pairs for 1D models with one degree of freedom.
We then used these families to construct Lax pairs
for the 1D systems discussed in Section \ref{sec:scla1D}, in cases where the chosen degrees of freedom are decoupled. The 1D systems
discussed in Section \ref{sec:scla1D} have $n+2$ degrees of freedom, 
$n$ of which correspond to $n$ scalar fields in four dimensions. Since in these decoupling cases, the analytic Lax pair description was based on $2(n + 2) \times 2 (n +2)$ matrices, for the ML experiments in the general case, we have taken the Lax pair for these systems to also have the same matrix dimensionality. 

\providecommand{\href}[2]{#2}\begingroup\raggedright\endgroup


\begin{thebibliography}{10}

\bibitem{Belinsky:1971nt}
V.A.~Belinsky and V.E.~Zakharov, \emph{{Integration of the Einstein Equations by the Inverse Scattering Problem Technique and the Calculation of the Exact Soliton Solutions}}, {\emph{Sov. Phys. JETP} {\bfseries 48} (1978) 985}.

\bibitem{Breitenlohner:1986um}
P.~Breitenlohner and D.~Maison, \emph{{On the Geroch Group}}, {\emph{Ann. Inst. H. Poincare Phys. Theor.} {\bfseries 46} (1987) 215}.

\bibitem{Nicolai:1991tt}
H.~Nicolai, \emph{{Two-dimensional gravities and supergravities as integrable system}}, \href{https://doi.org/10.1007/3-540-54978-1_12}{\emph{Lect. Notes Phys.} {\bfseries 396} (1991) 231}.

\bibitem{Lu:2007jc}
H.~Lu, M.J.~Perry and C.N.~Pope, \emph{{Infinite-dimensional symmetries of two-dimensional coset models coupled to gravity}}, \href{https://doi.org/10.1016/j.nuclphysb.2008.07.035}{\emph{Nucl. Phys. B} {\bfseries 806} (2009) 656} [\href{https://arxiv.org/pdf/0712.0615}{{\ttfamily 0712.0615}}].

\bibitem{Katsimpouri:2012ky}
D.~Katsimpouri, A.~Kleinschmidt and A.~Virmani, \emph{{Inverse Scattering and the Geroch Group}}, \href{https://doi.org/10.1007/JHEP02(2013)011}{\emph{JHEP} {\bfseries 02} (2013) 011} [\href{https://arxiv.org/pdf/1211.3044}{{\ttfamily 1211.3044}}].

\bibitem{Katsimpouri:2013wka}
D.~Katsimpouri, A.~Kleinschmidt and A.~Virmani, \emph{{An inverse scattering formalism for STU supergravity}}, \href{https://doi.org/10.1007/JHEP03(2014)101}{\emph{JHEP} {\bfseries 03} (2014) 101} [\href{https://arxiv.org/pdf/1311.7018}{{\ttfamily 1311.7018}}].

\bibitem{Chakrabarty:2014ora}
B.~Chakrabarty and A.~Virmani, \emph{{Geroch Group Description of Black Holes}}, \href{https://doi.org/10.1007/JHEP11(2014)068}{\emph{JHEP} {\bfseries 11} (2014) 068} [\href{https://arxiv.org/pdf/1408.0875}{{\ttfamily 1408.0875}}].

\bibitem{Camara:2017hez}
M.C.~C\^amara, G.L.~Cardoso, T.~Mohaupt and S.~Nampuri, \emph{{A Riemann-Hilbert approach to rotating attractors}}, \href{https://doi.org/10.1007/JHEP06(2017)123}{\emph{JHEP} {\bfseries 06} (2017) 123} [\href{https://arxiv.org/pdf/1703.10366}{{\ttfamily 1703.10366}}].

\bibitem{Cardoso:2017cgi}
G.L.~Cardoso and J.C.~Serra, \emph{{New gravitational solutions via a Riemann-Hilbert approach}}, \href{https://doi.org/10.1007/JHEP03(2018)080}{\emph{JHEP} {\bfseries 03} (2018) 080} [\href{https://arxiv.org/pdf/1711.01113}{{\ttfamily 1711.01113}}].

\bibitem{Aniceto:2019rhg}
P.~Aniceto, M.C.~C\^amara, G.L.~Cardoso and M.~Rossell\'o, \emph{{Weyl metrics and Wiener-Hopf factorization}}, \href{https://doi.org/10.1007/JHEP05(2020)124}{\emph{JHEP} {\bfseries 05} (2020) 124} [\href{https://arxiv.org/pdf/1910.10632}{{\ttfamily 1910.10632}}].

\bibitem{Penna:2021kua}
R.F.~Penna, \emph{{Einstein\textendash{}Rosen waves and the Geroch group}}, \href{https://doi.org/10.1063/5.0061929}{\emph{J. Math. Phys.} {\bfseries 62} (2021) 082503} [\href{https://arxiv.org/pdf/2106.13252}{{\ttfamily 2106.13252}}].


\bibitem{Camara:2022gvc}
M.C.~C\^amara and G.L.~Cardoso, \emph{{Generating new gravitational solutions by matrix multiplication}},  
{\em Proceedings of the Royal Society A:
  Mathematical, Physical and Engineering Sciences} {\bf 480} (2024), no.~2285,
\href{https://doi.org/10.1098/rspa.2023.0857}{{\tt
  https://doi.org/10.1098/rspa.2023.0857}},
\href{https://arxiv.org/pdf/2211.01702.pdf}{{\tt 2211.01702}}.



\bibitem{Camara:2024ham}
M.C.~C\^amara and G.L.~Cardoso, \emph{{Riemann-Hilbert problems, Toeplitz operators and ergosurfaces}},  
\href{https://doi.org/10.1007/JHEP06(2024)027}{\emph{JHEP} {\bfseries 06} (2024) 027}
[\href{https://arxiv.org/pdf/2404.03373}{{\ttfamily 2404.03373}}].



\bibitem{Figueras:2009mc}
P.~Figueras, E.~Jamsin, J.V.~Rocha and A.~Virmani, \emph{{Integrability of Five Dimensional Minimal Supergravity and Charged Rotating Black Holes}}, \href{https://doi.org/10.1088/0264-9381/27/13/135011}{\emph{Class. Quant. Grav.} {\bfseries 27} (2010) 135011} [\href{https://arxiv.org/pdf/0912.3199}{{\ttfamily 0912.3199}}].

\bibitem{Leigh:2014dja}
R.G.~Leigh, A.C.~Petkou, P.M.~Petropoulos and P.K.~Tripathy, \emph{{The Geroch group in Einstein spaces}}, \href{https://doi.org/10.1088/0264-9381/31/22/225006}{\emph{Class. Quant. Grav.} {\bfseries 31} (2014) 225006} [\href{https://arxiv.org/pdf/1403.6511}{{\ttfamily 1403.6511}}].

\bibitem{Klemm:2015uba}
D.~Klemm, M.~Nozawa and M.~Rabbiosi, \emph{{On the integrability of Einstein\textendash{}Maxwell\textendash{}(A)dS gravity in the presence of Killing vectors}}, \href{https://doi.org/10.1088/0264-9381/32/20/205008}{\emph{Class. Quant. Grav.} {\bfseries 32} (2015) 205008} [\href{https://arxiv.org/pdf/1506.09017}{{\ttfamily 1506.09017}}].


\bibitem{cmn}
G.L.~Cardoso, S.~Mahapatra, and S.~Nagy, \emph{{Weyl-Lewis-Papapetrou coordinates, self-dual Yang-Mills equations and the single copy}}, \href{https://link.springer.com/article/10.1007/JHEP10(2024)030}{\emph{JHEP} {\bfseries 10} (2024) 030} 
[\href{https://arxiv.org/pdf/2407.14392}{{\ttfamily 2407.14392}}].



\bibitem{Krippendorf:2021lee}
S.~Krippendorf, D.~Lust and M.~Syvaeri, \emph{{Integrability Ex Machina}}, \href{https://doi.org/10.1002/prop.202100057}{\emph{Fortsch. Phys.} {\bfseries 69} (2021) 2100057} [\href{https://arxiv.org/pdf/2103.07475}{{\ttfamily 2103.07475}}].

\bibitem{Lal:2023dkj}
S.~Lal, S.~Majumder and E.~Sobko, \emph{{The R-mAtrIx Net}},  \href{https://arxiv.org/pdf/2304.07247}{{\ttfamily 2304.07247}}.

\bibitem{Ferrara:1997tw}
S.~Ferrara, G.W.~Gibbons and R.~Kallosh, \emph{{Black holes and critical points in moduli space}}, \href{https://doi.org/10.1016/S0550-3213(97)00324-6}{\emph{Nucl. Phys. B} {\bfseries 500} (1997) 75} [\href{https://arxiv.org/pdf/hep-th/9702103}{{\ttfamily hep-th/9702103}}].

\bibitem{Goldstein:2005hq}
K.~Goldstein, N.~Iizuka, R.P.~Jena and S.P.~Trivedi, \emph{{Non-supersymmetric attractors}}, \href{https://doi.org/10.1103/PhysRevD.72.124021}{\emph{Phys. Rev. D} {\bfseries 72} (2005) 124021} [\href{https://arxiv.org/pdf/hep-th/0507096}{{\ttfamily hep-th/0507096}}].

\bibitem{Goldstein:2005rr}
K.~Goldstein, R.P.~Jena, G.~Mandal and S.P.~Trivedi, \emph{{A C-function for non-supersymmetric attractors}}, \href{https://doi.org/10.1088/1126-6708/2006/02/053}{\emph{JHEP} {\bfseries 02} (2006) 053} [\href{https://arxiv.org/pdf/hep-th/0512138}{{\ttfamily hep-th/0512138}}].

\bibitem{Goldstein:2014gta}
K.~Goldstein, V.~Jejjala and S.~Nampuri, \emph{{Hot Attractors}}, \href{https://doi.org/10.1007/JHEP01(2015)075}{\emph{JHEP} {\bfseries 01} (2015) 075} [\href{https://arxiv.org/pdf/1410.3478}{{\ttfamily 1410.3478}}].

\bibitem{Barisch:2011ui}
S.~Barisch, G.L.~Cardoso, M.~Haack, S.~Nampuri and N.A.~Obers, \emph{{Nernst branes in gauged supergravity}}, \href{https://doi.org/10.1007/JHEP11(2011)090}{\emph{JHEP} {\bfseries 11} (2011) 090} [\href{https://arxiv.org/pdf/1108.0296}{{\ttfamily 1108.0296}}].

\bibitem{Ishikawa_2021}
F.~Ishikawa, H.~Suwa and S.~Todo, \emph{Neural network approach to construction of classical integrable systems}, \href{https://doi.org/10.7566/jpsj.90.093001}{\emph{Journal of the Physical Society of Japan} {\bfseries 90} (2021) 093001} [\href{https://arxiv.org/pdf/2103.00372}{{\ttfamily 2103.00372}}].



\bibitem{daigavane2022learning}
A.~Daigavane, A.~Kosmala, M.~Cranmer, T.~Smidt and S.~Ho, \emph{Learning integrable dynamics with action-angle networks},  \href{https://arxiv.org/pdf/2211.15338}{{\ttfamily 2211.15338}}.

\bibitem{FUNAHASHI1989183}
K.-I.~Funahashi, \emph{On the approximate realization of continuous mappings by neural networks}, \href{https://doi.org/https://doi.org/10.1016/0893-6080(89)90003-8}{\emph{Neural Networks} {\bfseries 2} (1989) 183}.

\bibitem{Cybenko1989}
G.~Cybenko, \emph{{Approximation by superpositions of a sigmoidal function}}, \href{https://doi.org/10.1007/BF02551274}{\emph{Mathematics of Control, Signals and Systems} {\bfseries 2} (1989) 303}.

\bibitem{HORNIK1989359}
K.~Hornik, M.~Stinchcombe and H.~White, \emph{Multilayer feedforward networks are universal approximators}, \href{https://doi.org/https://doi.org/10.1016/0893-6080(89)90020-8}{\emph{Neural Networks} {\bfseries 2} (1989) 359}.

\bibitem{tensorflow2015-whitepaper}
M.~Abadi, A.~Agarwal, P.~Barham, E.~Brevdo, Z.~Chen, C.~Citro et~al., \emph{{TensorFlow}: Large-scale machine learning on heterogeneous systems},  \href{https://arxiv.org/pdf/1603.04467}{{\ttfamily 1603.04467}}.


\bibitem{lrpaper}
K.~Nakamura, B.~Derbel, K.-J.~Won and B.-W.~Hong, \emph{Learning-rate annealing methods for deep neural networks}, \href{https://doi.org/10.3390/electronics10162029}{\emph{Electronics} {\bfseries 10} (2021) }.

\bibitem{cranmer2020discovering}
M.~Cranmer, A.~Sanchez-Gonzalez, P.~Battaglia, R.~Xu, K.~Cranmer, D.~Spergel et~al., \emph{Discovering symbolic models from deep learning with inductive biases},  \href{https://arxiv.org/pdf/2006.11287}{{\ttfamily 2006.11287}}.

\end{thebibliography}
\end{document}